\documentclass[aps,prb,twocolumn,longbibliography,floatfix,superscriptaddress]{revtex4-2}
\usepackage{color}
\usepackage{amsmath}
\usepackage{amssymb}
\usepackage{bbold}
\usepackage{graphicx}
\usepackage{natbib}
\usepackage{enumitem}
\usepackage[normalem]{ulem}
\usepackage{subfloat}
\usepackage[caption=false,singlelinecheck=off]{subfig}

\usepackage[colorlinks=true,linkcolor=blue,urlcolor=blue,citecolor=blue]{hyperref}
\hypersetup{colorlinks=true}
\usepackage[all]{hypcap}

\newcommand{\be}{\begin{equation}}
\newcommand{\ee}{\end{equation}}
\newcommand{\bea}{\begin{eqnarray}}
\newcommand{\eea}{\end{eqnarray}}

\begin{document}

\title{
Two-dimensional electron hydrodynamics in a random array\\ of impenetrable obstacles:\\
Magnetoresistivity, Hall viscosity, and the Landauer dipole
}
\author{I. V. Gornyi}
\altaffiliation{Also was at Ioffe Institute, 194021 St.~Petersburg, Russia during the main stages of the project.}
\affiliation{\mbox{Institute for Quantum Materials and Technologies, Karlsruhe Institute of Technology, 76021 Karlsruhe, Germany}}
\affiliation{\mbox{Institut f\"ur Theorie der Kondensierten Materie, Karlsruhe Institute of Technology, 76128 Karlsruhe, Germany}}
\author{D. G. Polyakov}
\affiliation{\mbox{Institute for Quantum Materials and Technologies, Karlsruhe Institute of Technology, 76021 Karlsruhe, Germany}}

\begin{abstract}
We formulate a general framework to study the flow of the electron liquid in two dimensions past a random array of impenetrable obstacles in the presence of a magnetic field. We derive a linear-response formula for the resistivity tensor $\hat\rho$ in hydrodynamics with obstacles, which expresses $\hat\rho$ in terms of the vorticity and its harmonic conjugate, both on the boundary of obstacles. In the limit of rare obstacles, in which we calculate $\hat\rho$, the contributions of the flow-induced electric field to the dissipative resistivity from the area covered by the liquid and the area inside obstacles are shown to be equal to each other. We demonstrate that the averaged electric fields outside and inside obstacles are rotated by Hall viscosity from the direction of flow. For the diffusive boundary condition on the obstacles, this effect exactly cancels in $\hat\rho$. By contrast, for the specular boundary condition, the total electric field is rotated by Hall viscosity, which means the emergence of a Hall-viscosity-induced effective---proportional to the obstacle density---magnetic field. Its effect on the Hall resistivity is particularly notable in that it leads to a deviation of the Hall constant from its universal value. We show that the applied magnetic field enhances hydrodynamic lubrication, giving rise to a strong negative magnetoresistance. We combine the hydrodynamic and electrostatic perspectives by discussing the distribution of charges that create the flow-induced electric field around obstacles. We provide a connection between the tensor $\hat\rho$ and the disorder-averaged electric dipole induced by viscosity at the obstacle. This establishes a conceptual link between the resistivity in hydrodynamics with obstacles and the notion of the Landauer dipole. We show that the viscosity-induced dipole is rotated from the direction of flow by Hall viscosity.
\end{abstract}

\maketitle

\section{Introduction}
\label{s1}

Hydrodynamics of the electron liquid in solids has received renewed attention---after the early theoretical advances made about half a century ago \cite{gurzhi64,gurzhi68}---largely due to the recent surge of interest in the transport properties of {\it clean} two-dimensional (2D) electron systems. From the experimental perspective, the relatively clean electron systems of interest primarily include those in
high-mobility semiconductor structures \cite{molenkamp94,dejong95,levin18,gusev18,gusev20,gupta21a,keser21,wang22}, undoped graphene \cite{bandurin16,crossno16,kumar17,bandurin18,berdyugin19,sulpizio19,ku20,geurs20}, and pure quasi-2D (semi)metals \cite{moll16,*bachmann22,gooth17,vool21,aharon-steinberg22}.

In the context of electrons in the solid-state environment, conventional wisdom suggests that the hydrodynamic approach is accurate in describing viscous electron flows on spatial scales larger than the momentum-{\it density} relaxation length $l_{\rm ee}$ (resulting from total-momentum conserving electron-electron collisions) and smaller than the total-momentum relaxation length $l_{\rm P}$ (resulting from external perturbations, say, impurity-induced disorder or thermal lattice excitations). From the point of view of hydrodynamics, it is this separation of scales that distinguishes the exceptionally clean electron systems from more typical conductors. More specifically, it is the possibility of having a ``window" for the temperature $T$, as $T$ is varied, within which $l_{\rm ee}\alt l_{\rm P}$ even for low $T$, where the total-momentum conservation is broken by impurities or other static imperfections of the kind.

The advent of electron hydrodynamics in solids has substantially modified the conceptual framework by which to categorize the transport phenomena in interacting electron systems in condensed-matter physics. In the hydrodynamic limit ($l_{\rm ee}\ll l_{\rm P}$), the key notion to characterize collective electron transport becomes that of viscosity. The collective variable that is fundamental to the notion of viscosity is the space-time dependent drift velocity ${\bf v}({\bf r},t)$. The viscosity-based approach aims at describing the universal properties of an electron liquid that follow solely from the total charge, momentum, and energy conservation in the presence of friction between parts of the system which move, at given $t$, with different ${\bf v}({\bf r},t)$. In this approach, ${\bf v}({\bf r},t)$ obeys a continuity equation for the momentum density in the form of the Navier-Stokes equation \cite{lanlif6}, where friction is parametrized by the viscosity coefficients.

From the microscopic point of view, the viscous-flow description of an electron liquid hinges on the assumption that the system at any $\bf r$ is close to a local ($\bf r$-dependent) thermal equilibrium, established as a result of fast electron-electron collisions, in the moving [with the velocity ${\bf v}({\bf r},t)$] frame. Dissipation produced by viscosity is a measure of deviation from the local equilibrium. From a more general perspective on the collective response, the viscous hydrodynamics emerges as the limiting case of the theoretical framework that bridges the gap between the hydrodynamics and the collisionless (Vlasov) dynamics in Fermi liquids \cite{conti99,tokatly99,*tokatly00}. The microscopic consideration is also instrumental in representing the viscosity coefficients as Kubo formulas in terms of the correlation functions of the stress tensor, both in the classical and quantum formulations \cite{resibois77,mclennan60,bradlyn12,principi16}. This also indicates, in the case of charged particles, an inherent link between viscosity and the momentum dispersion of the conductivity (or resistivity) tensor \cite{resibois77,bradlyn12,hoyos12}.

Our purpose here is to formulate a framework to study hydrodynamic transport of the 2D electron liquid in a random ensemble of impenetrable obstacles and apply it to calculate the resistivity tensor $\hat\rho$ in the presence of a magnetic field. In particular, we will derive a general linear-response formula for $\hat\rho$ in hydrodynamics with obstacles. As a fundamentally compact relation, this formula expresses $\hat\rho$ in terms of the vorticity and its harmonic conjugate on the boundary of obstacles. The essential meaning of this formulation is that $\hat\rho$, defined by the average electric field induced by the electron flow, has two contributions, one of which comes from the area covered by the liquid and the other from the area inside obstacles. Remarkably, in the limit of rare obstacles, in which we calculate $\hat\rho$, the electric fields outside and inside obstacles contribute equally to the dissipative resistivity $\rho_{xx}$. These two contributions to $\rho_{xx}$ are associated, in a one-to-one correspondence, with the effect of viscous stress (electric fields outside obstacles) and pressure exerted by obstacles on the liquid (fields inside them).

We will particularly focus on the role played by Hall viscosity in transport of the electron liquid past obstacles. We will show that the averaged flow-induced electric fields outside and inside obstacles are rotated by Hall viscosity with respect to the direction of the averaged velocity. With regard to this effect of Hall viscosity, the boundary conditions imposed on the flow on the boundaries of obstacles become a matter of major importance. For the diffusive boundary condition, the Hall-viscosity-induced modifications of the electric fields outside and inside obstacles exactly compensate each other in the space-averaged total electric field, so that $\hat\rho$ is independent of Hall viscosity.  By contrast, for the specular boundary condition, the total electric field is modified by Hall viscosity.

A conceptually significant point that follows from this consideration is the emergence of a Hall-viscosity-induced effective magnetic field in the flow perturbed by obstacles for a finite degree of ``specularity" in the boundary condition. Its effect on the Hall resistivity $\rho_{xy}$ is particularly prominent in that it modifies the Hall constant compared to the universal value characteristic of the Drude formula (and of the result for $\rho_{xy}$ in the case of the diffusive boundary condition for that matter). Within a more conventional hydrodynamic context, we will also calculate the drag and lift forces exerted by the electron liquid on the obstacle, where the lift force emerges entirely because of Hall viscosity and counterbalances the force exerted on the liquid by the effective magnetic field.

Apart from the Hall-viscosity-induced modification of $\hat\rho$, we will provide a controlled description of another effect the external magnetic field $B$ has on $\hat\rho$ in the hydrodynamic regime, namely the enhancement of hydrodynamic lubrication in the flow of charge through an array of obstacles, with $\rho_{xx}$ vanishing to zero in the limit $B\to\infty$. This effect is in stark contrast to the Drude noninteracting regime (characterized, from the point of view of relaxation processes, solely by $l_{\rm P}$). In light of this picture, the magnetic-field induced lubrication makes hydrodynamic transport accessible and detectable in the measurement of the bulk resistivity in samples that need not be narrow.

We will round out our consideration of hydrodynamic transport in a random obstacle array with a calculation of the spatial distribution of charges that create the flow-induced electric field around obstacles. This will relate the resistivity in the hydrodynamic regime with the disorder-averaged electric dipole induced by viscosity at the obstacle, thus establishing a conceptual perspective which links hydrodynamics in disordered media to the notion of the Landauer dipole \cite{landauer57,*landauer75}. In particular, we will show that the viscosity-induced dipole is rotated from the direction of flow by an angle dependent on the ratio of the Hall and dissipative viscosity coefficients.

Methodologically, we will formulate a model to explore the electron flow around a single obstacle and solve it in detail for the case when both the dissipative and Hall viscosities are present. This solution will be used to perform disorder averaging up to the leading terms in $\hat\rho$ induced by Hall viscosity. We will complement this approach by providing a mean-field solution of the hydrodynamic problem in a random array of obstacles.

The paper is organized as follows. In Sec.~\ref{s2}, we present background material concerning viscosity in the presence of a magnetic field. In Sec.~\ref{s3}, we provide a general perspective as to the compressibility of the 2D electron liquid in the hydrodynamic formalism, particularly with regard to the charges created by the flow around obstacles. In Sec.~\ref{s4}, we discuss a general picture of how the magnetic field affects hydrodynamics of the electron liquid. In Sec.~\ref{s5}, we write the boundary conditions for the flow at $B\neq 0$. In Sec.~\ref{s6a}, we formulate the general framework for calculating the resistivity in the hydrodynamic problem. In Sec.~\ref{s7}, we study the flow past a single obstacle in the presence of both the dissipative and Hall viscosities. In Sec.~\ref{s9}, we perform disorder averaging and calculate the magnetoresistivity tensor for the hydrodynamic flow through the array of obstacles. In Sec.~\ref{s8}, we consider the charge distribution induced by the flow and its dependence on the magnetic field. In Sec.~\ref{s10}, we critically discuss some of the experimental results. Section \ref{s11} provides a summary. In Appendix \ref{a}, we consider the mean-field formulation of transport in the obstacle array. Appendix \ref{b} extends the discussion of the flow-induced charge distribution in Sec.~\ref{s8}.

\section{Viscosity of a magnetized plasma}
\label{s2}

As a general method and ideology, hydrodynamics of viscous {\it conducting} (thus subject to the Lorentz force) fluids has been actively investigated over the past sixty or so years---primarily in the context of an electron-ion plasma,
both in the Galilean-invariant and relativistic limits, with emphasis on the viscous properties of a magnetized plasma \cite{lanlif10} (also in connection with magnetohydrodynamics, where fluid mechanics is fundamentally coupled to electromagnetism). It has been well understood that viscosity is modified by a magnetic field \cite{chapman70,steinberg58,kaufman60,braginskii65,*braginskii58,alekseev16} (see also Ref.~\cite{lanlif10} for a general discussion illustrated by the limit of large $B$).

For the case of an incompressible liquid (considered in the present paper, with a disclaimer specified in Sec.~\ref{s3}), the two kinematic viscosity coefficients $\nu$ and $\nu_{\rm H}$ that describe purely transverse (with respect to the direction of the magnetic field) deformations are given by \cite{chapman70,steinberg58,kaufman60,braginskii65,*braginskii58,alekseev16,2vs3}
\be
\nu=\nu_0\,\frac{1}{1+(2\omega_c\tau_{\rm ee})^2}~,\quad \nu_{\rm H}=\nu_0\,\frac{2\omega_c\tau_{\rm ee}}{1+(2\omega_c\tau_{\rm ee})^2}~,
\label{1}
\ee
where $\omega_c>0$ \cite{sign_hallvisc} is the cyclotron frequency, $\tau_{\rm ee}=l_{\rm ee}/v_F$ with $v_F$ being the Fermi velocity (assuming the degenerate case of low $T$), $\nu_0=v_F^2\tau_{\rm ee}/4$, and the magnetohydrodynamic effects \cite{braginskii65}, normally weak in the solid-state context, are neglected.
Throughout the paper, viscosity that stems from electron scattering off static disorder is neglected, under the assumption that $1/\tau_{\rm ee}$ is much larger than the disorder-induced relaxation rate of the second angular harmonic of the distribution function
(for disorder-induced viscosity, see Refs.~\cite{burmistrov19,zakharov21}).

One important caveat to note is that, in the 2D case, the relaxation rates for the even and odd (in momentum space) parts of the electron distribution function that are induced by electron-electron scattering are generically vastly different in the limit of low $T$ \cite{gurzhi95,ledwith17,ledwith19,ledwith19a,alekseev20,kryhin23,hofmann23,kryhin23a}, which may lead to the emergence of a ``quasihydrodynamic" regime \cite{gurzhi95,ledwith17,ledwith19,hofmann22,kryhin23a} not characterizable by a single electron-electron scattering time. In the present paper, however, we consider the ``orthodox" hydrodynamic regime, in which $\tau_{\rm ee}$ is the relaxation time for the second angular harmonic of the distribution function.

The magnetic field is seen to manifest itself in Eqs.~(\ref{1}) in a twofold manner: as $B$ increases, the ``conventional" viscosity coefficient $\nu$ decreases, down to zero in the limit $B\to\infty$, and there emerges the Hall viscosity coefficient $\nu_{\rm H}$. The contribution of $\nu_{\rm H}$ to viscous heating is zero (``this stress is `orthogonal' to the strain," as it was elegantly formulated in Ref.~\cite{kaufman60}), i.e., $\nu_{\rm H}$ for arbitrary $\omega_c\tau_{\rm ee}$ describes, in contrast to $\nu$, nondissipative transport (similarly to the Hall resistivity $\rho_{xy}$). The nonvanishing of $\nu_{\rm H}$ in the frictionless limit of $\tau_{\rm ee}\to\infty$ for given $\omega_c$ in Eq.~(\ref{1}) can be rationalized either within the kinetic approach or, equivalently, from the perspective of the geometric interpretation of viscosity of noninteracting electrons in Landau levels through the response to a variation of the metric tensor \cite{avron95,read11,abanov14,hoyos14}.


Apart from the experimental works \cite{molenkamp94,dejong95,levin18,gusev18,gusev20,gupta21a,keser21,wang22,bandurin16,crossno16,kumar17,bandurin18,berdyugin19,sulpizio19,ku20,
geurs20,moll16,gooth17,vool21,aharon-steinberg22}, electrically measurable manifestations of viscosity in dc transport in specific setups of 2D electron devices were discussed in
Refs.~\cite{tomadin14,torre15,scaffidi17,levitov16,pellegrino16,guo17,falkovich17,delacretaz17,pellegrino17,holder19},
with Refs.~\cite{scaffidi17,delacretaz17,pellegrino17,holder19} focusing on Hall viscosity. Viscous transport in undoped graphene \cite{mueller09} shows a number of peculiarities associated with the linear dispersion relation of massless Dirac fermions near the crossing point and the presence of both electron and hole liquids, for a review see Refs.~\cite{narozhny17,lucas18a,narozhny19,*narozhny22}. Here, we focus on viscous hydrodynamics of massive electrons.

\section{``Incompressible" electron liquid in two dimensions}
\label{s3}

For the 2D incompressible electron liquid (defined by assuming a constant electron density $n$, the meaning of the quotation marks in the title
will become clear shortly) with the mass and charge densities $mn$ and $-en$ (with $e>0$), respectively, the continuity equation and the linearized Navier-Stokes equation with the viscosity coefficients for $B\neq 0$ from Eq.~(\ref{1}) are written as
\begin{align}
\nabla{\bf v}&=0~,\label{2}\\
\partial_t{\bf v}&=\nabla\phi-\omega_c({\bf v}\times{\bf n})+\nu\nabla^2{\bf v}-\nu_{\rm H}(\nabla^2{\bf v}\times{\bf n})~,
\label{3}
\end{align}
where $\phi$ in the pressure term is related to the electric potential $V$ by
\be
\phi=\frac{e}{m}V~,
\label{3h}
\ee
and $\bf n$ is the unit vector in the direction of the (perpendicular) magnetic field. Note that, within the hydrodynamic description of the incompressible liquid of {\it charged} particles (``plasma") at a homogeneous entropy density (we neglect throughout the paper the contributions to pressure produced by flow-induced inhomogeneities of both the chemical potential and temperature), the gradient $-mn\nabla\phi$ of the flow-induced pressure is solely due to long-range electric forces. For the degenerate electron liquid, this implies neglecting the force that comes from a spatial variation of the degeneracy pressure $(\pi\hbar^2/m)n^2$ (per spin). Without much discussion of the origin of the electric forces in terms of charges, this fact was used in Ref.~\cite{levitov16} to demonstrate the essential difference between the electric potential profiles generated by the viscous and Ohmic incompressible flows. An important basic question, worth answering here, is about what charges produce the electric forces \cite{compress}. This question is not exactly trivial, as we discuss next.

A conceptually significant point, which does not seem to have been generally appreciated in the literature, is that the viscous electron liquid obeying Eqs.~(\ref{2}) and (\ref{3}) in the 2D case or their direct analog in the three-dimensional (3D) case \cite{three_d} is charge-neutral locally (i.e., incompressible as such) {\it only} in the 3D case (and even then generically only for $B=0$). For a 2D liquid, the situation is qualitatively different in that the solution to Eqs.~(\ref{2}) and (\ref{3}) for a steady-state viscous flow past hard-wall scatterers (which are encoded in the boundary conditions to these equations) is necessarily associated with a generation of charges in the {\it bulk} of the liquid (even for $B=0$). This is despite the by now common designation for Eqs.~(\ref{2}) and (\ref{3}) as an incompressible model.

Consider first the case of $B=0$. It is instructive to contrast the 2D and 3D geometries. In the 3D case, the electric field ${\bf E}=-(m/e)\nabla\phi$ is produced at $B=0$ by charges that are ``external" within the formalism of Eqs.~(\ref{2}) and (\ref{3}), in the sense that the Poisson equation reads
\be
{\rm 3D}_{B=0}\,{\rm :}\qquad \nabla^2\phi=0~,\quad n=n_0
\label{3a}
\ee
everywhere {\it inside} the liquid [as can be seen by applying $\nabla$ to Eq.~(\ref{3})],
where $n_0$ is the equilibrium density (homogeneous between the scatterers).
That is, the viscosity-induced field in the presence of a flow is produced by charges that sit {\it exactly} on the boundaries of the liquid. In the bulk, the liquid can be thought of as being exactly incompressible.

It is a subtle feature of the hydrodynamic description that the very existence of a nonequilibrium solution to Eqs.~(\ref{2}) and (\ref{3}) for $\nu\neq 0$ is entirely \cite{poiseuille} due to a tacitly assumed nonzero compressibility of the electron liquid in the vicinity of the boundary. Indeed, it is only because of a compressibility that charges can be generated on the boundary when the liquid is pushed towards or away from it (assuming, as we do here, that the shape of the boundary is rigidly fixed, which is the case for a hard wall). The beauty of the hydrodynamic formalism in the incompressible limit [defined by Eq.~(\ref{2})] is that the field $\nabla\phi$ is viewed as freely adjusting itself to the flow constrained solely by the boundary conditions imposed on the field ${\bf v}$, so that the value of the ``boundary" compressibility completely drops out from the distribution of ${\bf v}$.

In a 2D liquid, the equation $\nabla^2\phi=0$ holds at $B=0$ as well, with $\nabla$ acting within the plane, as follows from Eq.~(\ref{3}) in similarity with the 3D case. By contrast, however, the flow in a viscous 2D liquid, obeying Eqs.~(\ref{2}) and (\ref{3}), in the presence of impenetrable scatterers generates charges that are now spread over the bulk of the liquid \cite{edge}. The flow-induced charge density is smooth and falls off away from the boundaries in a power-law manner, namely
\be
{\rm 2D}_{B=0}\,{\rm :}\qquad \nabla^2\phi=0~,\quad n-n_0\propto 1/r^2~,
\label{3d2}
\ee
where $r$ is the distance to the scatterer, as we will demonstrate below in Sec.~\ref{s8}. It is worth emphasizing that the gradient of the electric potential that emerges to counterbalance the viscous forces {\it necessitates} the production of an inhomogeneous charge density in the 2D flow while the gradient of the chemical potential \cite{compress}, associated with the charge inhomogeneity, may be {\it totally} neglected in the balance of forces in Eq.~(\ref{3}).

Because the gradient of the chemical potential does not explicitly enter Eq.~(\ref{3}), the compressibility itself does not explicitly show up in the solution for ${\bf v}$ and in the corresponding profile of $\phi$, similarly to the 3D case. It is the latter circumstance that arguably justifies the use of the term ``incompressible" with regard to the model of Eqs.~(\ref{2}) and (\ref{3}) in a 2D liquid---but only with the caveat specified above [``inherently" finite modulation of $n$, Eq.~(\ref{3d2})]; hence the quotation marks in the title of this section. As a matter of fact, Eqs.~(\ref{2}) and (\ref{3}), when applied to the electron liquid, may describe the limit of perfect screening (``infinitely {\it high} compressibility"), where $-(m/e)\nabla\phi$ is the ``residual" (screened) electric field \cite{compress}. This is precisely the limit we consider when discussing the density profile in Sec.~\ref{s8}.

A useful way to account for the emergence of the bulk 2D charges is to realize that the flow ${\bf v}({\bf r})$ in the ${\bf r}=(x,y)$ plane in Eq.~(\ref{3}) in the presence of impenetrable disks is exactly the same as in an incompressible 3D liquid flowing past impenetrable cylinders obtained by ``translating" the disks along the $z$ axis. Therefore, the pressure profiles are also the same in the $(x,y)$ plane, i.e., the electric field $-(m/e)\nabla\phi_{\rm disk}(x,y)$ created by a disk in the 2D liquid is the same as the $z$ independent electric field $-(m/e)\nabla\phi_{\rm cyl}(x,y)$ created by a cylinder in the 3D liquid:
\be
\nabla\phi_{\rm disk}(x,y)=\nabla\phi_{\rm cyl}(x,y)~.
\label{3d1}
\ee
We will use this identity in Sec.~\ref{s8}.

The difference between the 2D and 3D cases is in the distribution of charges that create $\nabla\phi_{\rm disk}(x,y)$ and $\nabla\phi_{\rm cyl}(x,y)$, respectively. In the 3D case, all charges sit on the surface of the cylinder [Eq.~(\ref{3a})]. Therefore, to produce the same electric field in the plane of a 2D liquid, the 2D charge density must necessarily be finite away from the disk. More specifically, the interplay of 2D hydrodynamics and 3D electrostatics in a 2D liquid produces ${\bf E}({\bf r})$ at $B=0$ that satisfies the Poisson equation (inside the liquid and on its two surfaces, with a 2D vector $\nabla$) of the form
\be
\nabla {\bf E}\vert_{z\to 0}=0~,\quad \partial_zE_z\vert_{z\to 0}=-4\pi e(n-n_0)\delta(z)~.
\label{3b1}
\ee
The profile of $E_{x,y}$ in Eq.~(\ref{3b1}) is peculiar in that the \mbox{in-plane} field is incompressible (in addition to being irrotational, so that altogether the 2D electric field is a harmonic function inside the liquid) while the charge density $n-n_0$ is inhomogeneous in the plane.

We have thus seen that the 2D and 3D cases for $B=0$ differ in an essential manner in Eqs.~(\ref{3a}) and (\ref{3d2}). If $B\neq 0$, however, $\nabla^2\phi$ becomes nonzero and charges are generically induced in the bulk of a flow irrespective of dimensionality. In a 2D liquid, described by Eqs.~(\ref{2}) and (\ref{3}), the expression for $\nabla^2\phi$ in terms of $\bf v$ takes a particularly simple form in the static limit:
\be
\nabla^2\phi=s\omega_c\Omega~,
\label{11a}
\ee
where
\be
\Omega=(\nabla\times{\bf v}){\bf e}_z
\label{3f}
\ee
is the 2D (scalar) vorticity, ${\bf e}_z$ is the unit vector in the $z$ direction, and
\be
s={\bf n}{\bf e}_z
\label{3f2}
\ee
is equal to $\pm 1$ depending on the orientation of the magnetic field. Note that, in Eq.~(\ref{11a}), the Hall viscosity term in the Navier-Stokes equation does not contribute to $\nabla^2\phi$ [it does only if $\Omega$ depends on $t$, in which case the right-hand side of Eq.~(\ref{11a}) has one more term, namely $s\nu_{\rm H}\nabla^2\Omega$, which is then nonzero because of $\nu\nabla^2\Omega=\partial_t\Omega$].

In a 3D liquid, $\nabla^2\phi$ in the static limit contains not only a contribution from the Lorentz force term, the same as in Eq.~(\ref{11a}), but also a contribution from the viscosity terms, which is only nonzero, in view of Eq.~(\ref{3a}), because of the {\it anisotropic} (transverse vs.\ longitudinal with respect to the magnetic field) modification \cite{lanlif10,chapman70,kaufman60,braginskii65,*braginskii58} of the viscosity tensor by $B\neq 0$. For the 3D flow, a nonzero $\nabla^2\phi$ is directly linked to the production of charges. In a 2D liquid, as mentioned below Eq.~(\ref{3b1}), the relation between $\nabla^2\phi$ and $n$ (with $\nabla$ acting in the plane) is more subtle: $\omega_c\neq 0$ means a generation of {\it additional}, compared to the case of zero $B$, charges in the bulk. We will discuss the distribution of $\phi$ and $n$ in more detail in Sec.~\ref{s8}.

\section{Hydrodynamic velocity in a magnetic field}
\label{s4}

Having highlighted in Sec.~\ref{s3} the essential difference between the 2D and 3D hydrodynamics of a plasma with regard to the distribution of $n$ in the bulk of a flow and specified precisely in what sense Eqs.~(\ref{2}) and (\ref{3}) can be thought of describing an ``incompressible" liquid,
we
focus below on the 2D case. We now turn to the role the magnetic field-induced quantities $\omega_c$ and $\nu_H$ play in the distribution of $\bf v$ in Eqs.~(\ref{2}) and (\ref{3}).

Note that the Hall viscosity term in Eq.~(\ref{3}), rewritten with the use of Eq.~(\ref{2}) as the force
\be
{\bf F}^{\rm H}_v=-m(s\nu_{\rm H})\nabla\Omega~,
\label{3b}
\ee
is associated with the vorticity (\ref{3f}), with $-s\nu_H\Omega$ playing in Eq.~(\ref{3b}) the role of a ``potential" additional to $\phi$ in Eq.~(\ref{3}). Apart from modifying $\nu$, the magnetic field is thus seen to produce two distinctly different forces in Eq.~(\ref{3}): the Lorentz (cyclotron) force ${\bf F}_c=-m\omega_c({\bf v}\times{\bf n})$, which acts ``directly" on $\bf v$, and the transverse drag force ${\bf F}^{\rm H}_v$ due to the spatial variation of $\bf v$, which acts on the gradient of $\Omega$. The force ${\bf F}^{\rm H}_v$ is generically much stronger than the dissipative viscosity force ${\bf F}_v=m\nu\nabla^2{\bf v}$ [expressible in terms of $\Omega$ as $-m\nu(\nabla\times\Omega{\bf e}_z)$] in the large-$B$ limit of $\nu_H\gg\nu$.

It is good to be clear about the interplay of the two forces in the charge flow past obstacles. Certain conclusions about the effect of the magnetic field on the distribution of ${\bf v}$ can be formulated in rather general terms. Both magnetic field-induced forces ${\bf F}_c$ and ${\bf F}^{\rm H}_v$ in Eq.~(\ref{3}) can be incorporated in a shift of the potential $\phi\to\phi_{\rm H}$,
where
\be
\phi_{\rm H}=\phi+s\!\left(\omega_c+\nu_{\rm H}\nabla^2\right)\!\psi
\label{3d}
\ee
is expressed through the stream function $\psi$, defined in the 2D incompressible liquid by
\be
{\bf v}=\nabla\times\psi{\bf e}_z
\label{3e}
\ee
and related to the vorticity by
\be
\Omega=-\nabla^2\psi~.
\label{3g}
\ee

As can be seen from Eq.~(\ref{3}), by acting on it with ${\bf n}\times\nabla$, within the space in which the total-momentum conservation is not broken, i.e., between the obstacles, $\psi$ obeys the biharmonic equation
\be
\nabla^4\psi=0
\label{11}
\ee
for $\partial_t\Omega=0$. Equation (\ref{11}) is a hallmark \cite{lanlif6} of a 2D incompressible flow in the stationary limit (considered throughout in this paper) if one neglects (``linear response") the quadratic-in-${\bf v}$ inertia term in the Navier-Stokes equation, as was assumed in Eq.~(\ref{3}). In view of Eq.~(\ref{3g}), another way to state Eq.~(\ref{11}) is that $\Omega$ is a harmonic function,
\be
\nabla^2\Omega=0
\label{11b}
\ee
[cf.\ the comment below Eq.~(\ref{3f2})].

According to Eq.~(\ref{3d}), the validity of Eq.~(\ref{11}) does not depend on the presence or absence of the magnetic field. It follows, importantly, that the influence of the (homogeneous) magnetic field on $\psi({\bf r})$ and, by means of Eq.~(\ref{3e}), on ${\bf v}({\bf r})$
is entirely encoded in the boundary conditions to Eq.~(\ref{3}) [imposed on ${\bf v}({\bf r})$ and, possibly, its derivatives on both the external boundary of the system and on the boundaries of obstacles in the bulk of the flow]. If these are $B$ independent, the magnetic field does not affect the flow. Without taking Hall viscosity into account, the independence of $\psi({\bf r})$ on $B$ was pointed out in Ref.~\cite{falkovich17}. As will be discussed in Sec.~\ref{s5}, the boundary conditions generically {\it depend} on $B$, namely through Hall viscosity. That is, the force ${\bf F}^{\rm H}_v$, which does not deflect the flow in the bulk locally, affects the flow through the boundary conditions.

Having made the general conclusion about the effect of a magnetic field on the flow, it is worth noting that---in the broader context of odd viscosity \cite{avron98}---the Navier-Stokes equation is often represented in the form that contains the Hall-viscosity term but not the Lorentz-force term (both generically allowed by broken time-reversal symmetry). On the one hand, this may be a matter of notation, because the Lorentz force can be understood as being incorporated in the pressure term: in the 2D case, by adding the stream function $\psi$ to the pressure [Eq.~(\ref{3d})]. On the other hand, one can think of hydrodynamics of ``parity-violating fluids" that possess Hall viscosity in the absence of any external magnetic field \cite{lapa14,lucas14,ganeshan17,banerjee17,lou22,fruchart23}. In our model, both the Lorentz and Hall-viscosity forces are ``part of the equation." With regard to Hall viscosity, however, 2D parity-violating fluids in the absence of a magnetic field, with active chiral fluids (``spinners") as a prominent example, and 2D magnetized electron liquids share much of the phenomenology.

\section{Boundary condition in a magnetized viscous liquid}
\label{s5}

We introduce disorder in the viscous liquid by adding randomly placed, with the density $n_d$, rare impenetrable obstacles (``voids"), each in the form of a disk of radius $R$. We assume that the interior of the voids is neutral, i.e., contains no background charges. The fact that rare---even ``pointlike," of radius $R\ll l_{\rm ee}$---obstacles can produce a hydrodynamic contribution to the {\it conductivity} that would dominate over the Drude contribution was pointed out in Ref.~\cite{hruska02}, emphasizing the effect of logarithmically singular long-range hydrodynamic correlations in the 2D case (``Stokes paradox") \cite{lamb11,lanlif6}. For $B=0$ (and the sticky boundary condition), the friction force exerted on a moving fluid by the circle-shaped hard obstacle in the hydrodynamic regime of $R\gg l_{\rm ee}$ is given by Stokes' formula \cite{lamb11,lanlif6}. As argued in Ref.~\cite{guo16},  for $R\ll l_{\rm ee}\ll n_d^{-1/2}$, not only the hydrodynamic correlations on spatial scales between $l_{\rm ee}$ and $n_d^{-1/2}$ lead to the logarithmic enhancement \cite{hruska02} of the conductivity, but also multiple collisions of a given electron with the hard obstacle on scales between $R$ and $l_{\rm ee}$ contribute to the logarithmic singularity. In the present paper, however, we only consider the hydrodynamic regime, assuming that the radius of the obstacles is larger than the spatial scale over which the local equilibrium is established \cite{more_gen}.

Scattering of the electron flow by a ``large-scale" hard obstacle is the most conventional, from the hydrodynamic perspective, type of a local perturbation---and it is our goal to explore the consequences of $B\neq 0$, especially in the presence of Hall viscosity, for the electrical resistivity of an electron liquid in this (historically, ``hydrodynamic" in the most straightforward sense) limit. Conceptually important differences arising in the hydrodynamic limit in the case of smooth weak disorder, as opposed to the case of rare strong scatterers, were formulated in Ref.~\cite{andreev11} (see also Ref.~\cite{levchenko17} for a hydrodynamic description of magnetotransport in the limit of smooth disorder; for a broader perspective, see Ref.~\cite{lucas18}).

In our model, the boundary conditions to Eqs.~(\ref{2}) and (\ref{3}) are to be fixed on the boundaries of the voids. Two types of the boundary conditions that we discuss here correspond to two limiting cases of diffusive and specular electron scattering on the boundaries. For a disk centered at ${\bf r}=0$, the diffusive (sticky, or ``no-slip" in the hydrodynamic context) condition reads
\be
{\bf v}({\bf r})\vert_{r=R}=0~.
\label{4}
\ee

The specular (``no-stress") condition requires that the normal component of the velocity $v_r({\bf r})$ and the nondiagonal [``radial-tangential," in the polar coordinates $(r,\varphi)$] component of the stress tensor vanish on the boundary, where the latter condition is equivalent, in view of the former, to the vanishing of the radial-tangential component $\Pi_{r\varphi}({\bf r})$ of the momentum flux density tensor (contributes $-\partial_\varphi\Pi_{r\varphi}/r$ to $\partial_tv_r$) \cite{bubble}. The specular condition thus reads
\be
v_r({\bf r})\vert_{r=R}=0~,\qquad\Pi_{r\varphi}({\bf r})\vert_{r=R}=0~.
\label{5}
\ee
In the incompressible liquid, the Hall component of $\Pi_{r\varphi}$ is given by $-2s\nu_{\rm H}\partial_rv_r$, so that the second condition in Eq.~(\ref{5}) is written as
\be
\nu\left(\partial_rv_\varphi+\frac{1}{r}\,\partial_\varphi v_r-\frac{1}{r}\,v_\varphi\right)+2s\nu_{\rm H}\partial_rv_r=0
\label{6}
\ee
for $r=R$. More general boundary conditions, using the notion of the Navier slip length, were discussed in
Refs.~\cite{torre15,levitov16,pellegrino16,pellegrino17,kiselev18} (in Refs.~\cite{pellegrino17,kiselev18} also within a kinetic equation approach).

Importantly, the boundary condition (\ref{6}) is affected by Hall viscosity \cite{pellegrino17} and thus necessarily, in our model, by the magnetic field, even if one neglects the dependence of the dissipative coefficient $\nu$ on $B$---see also an analogous modification of the boundary condition in the edge magnetoplasmon problem in Ref.~\cite{cohen18}. This is in contrast to Ref.~\cite{falkovich17}, which studied the effect of the Lorentz force ${\bf F}_c$ on viscous transport---but where neither Hall viscosity in the boundary condition [Eq.~(\ref{6})] nor the Hall viscosity force ${\bf F}^{\rm H}_v$ [Eq.~(\ref{3b})] inside the liquid were included in the calculation. Note also that the dependence of the {\it specular} boundary condition on $\nu_{\rm H}$ entails its dependence on the {\it dissipative} viscosity coefficient as well (which is trivial by dimensionality but demonstrates a nontrivial interplay of dissipative and nondissipative processes for specular scattering at $B\neq 0$).
Within the context of active chiral liquids, the boundary condition in the form of Eq.~(\ref{6}) arises \cite{lou22} if one sends the rotational viscosity coefficient to zero while keeping the odd viscosity coefficient finite.

It is worthwhile to mention that, in contrast to conventional (``molecular") viscous fluids, Eq.~(\ref{5}) appears to be much more adequate [compared to Eq.~(\ref{4})] to describe at least some of the electron systems that are experimentally relevant to the measurements of viscosity effects. For example, multiple geometric-resonance peaks observed in magnetic focusing experiments in GaAs moderate- and high-mobility heterostructures \cite{aidala07,gupta21} and graphene \cite{taychatanapat13,morikawa15} indicate that boundary scattering in these samples was to a large degree specular.

\section{Resistivity in hydrodynamics}
\label{s6a}

For an arbitrary stationary flow describable by Eqs.~(\ref{2}) and (\ref{3}), the potential $\phi$ that obeys Eq.~(\ref{3}) is given by a sum of three terms:
\be
\phi=-\nu\widetilde\Omega+s\left(\nu_{\rm H}\Omega-\omega_c\psi\right)~,
\label{40a}
\ee
where those in the brackets correspond to the magnetic-field-induced terms in Eq.~(\ref{3d}), and $\widetilde\Omega$ is the harmonic conjugate of $\Omega$, with $\widetilde\Omega$ and $\Omega$ related by
\be
\nabla\widetilde\Omega=-\nabla\times\Omega{\bf e}_z
\label{41b}
\ee
[which is the Cauchy-Riemann condition for the analytic function $\Omega(x,y)+i\widetilde\Omega(x,y)$ of the complex variable $x+iy$]. Equation (\ref{40a}) can also be represented in terms of the potential $\phi_{\rm H}$ from Eq.~(\ref{3d}) as $\phi_{\rm H}=-\nu\widetilde\Omega$.

\subsection{Average electric field}
\label{s6aA}

Green's theorem applied to an arbitrary area integral of the curl-free 2D field ${\bf E}=-\nabla V$ reads
\be
\int\!d^2{\bf r}\,{\bf E}={\bf e}_z\times \left(\oint_{\rm ext}-\oint_{\rm int}\right)d{\bf l}\,V~,
\label{n1}
\ee
where the contour integration is performed along the exterior (ext) and interior (int) boundaries of the area, in both cases (here and everywhere below) in the counterclockwise direction. In particular, for $V=(m/e)\phi$ from Eq.~(\ref{40a}), the relation (\ref{n1}) allows one to express the electric field averaged over the total area $S$ of the system, which includes empty space inside obstacles,
\be
\langle{\bf E}\rangle=\frac{1}{S}\int\!d^2{\bf r}\,{\bf E}~,
\label{n2}
\ee
in terms of the hydrodynamic variables $\widetilde\Omega$, $\Omega$, and $\psi$ on the boundary of the sample only, with only $\oint_{\rm ext}$ in Eq.~(\ref{n1}). Specifically,
\be
\langle{\bf E}\rangle=\langle{\bf E}_{\rm H}\rangle-\frac{m}{eS}\,{\bf e}_z\times\oint_{sb}\!d{\bf l}\,\left(\nu\widetilde\Omega-s\nu_{\rm H}\Omega\right)~,
\label{41j}
\ee
where
\be
\langle{\bf E}_{\rm H}\rangle=\frac{m}{e}s\omega_c\left({\bf e}_z\times\langle{\bf v}\rangle\right)
\label{41g}
\ee
is the Hall field counterbalancing the Lorentz force averaged over the liquid, with $\langle{\bf v}\rangle$ given by
\be
\langle{\bf v}\rangle=\frac{1}{S}\int\!d^2{\bf r}\,{\bf v}~,
\label{12e2}
\ee
and the contour of integration in $\oint_{sb}$ runs along the sample boundary. The Hall-field term (\ref{41g}) comes from the integral of $\psi$ along the boundary of the sample, because of the identity
\be
\frac{1}{S}\oint_{sb}d{\bf l}\,\psi=-\langle{\bf v}\rangle~.
\label{41f}
\ee

To calculate the average (\ref{n2}), it is, however, more convenient---and also more instructive---to split it into two parts:
\be
\langle{\bf E}\rangle=\langle{\bf E}\rangle_{\rm obs}+\langle{\bf E}\rangle_{\rm liq}~,
\label{n3}
\ee
where the former term is the contribution to the integral (\ref{n2}) of the area inside obstacles and the latter of the area covered by the liquid. Integrating over the interior of obstacles, the substitution of Eq.~(\ref{40a}) in Eq.~(\ref{n1}) gives
\be
\langle{\bf E}\rangle_{\rm obs}=-\frac{mN}{eS}\,{\bf e}_z\times\left\langle\oint_{ob}\!d{\bf l}\,\left(\nu\widetilde\Omega-s\nu_{\rm H}\Omega\right)\right\rangle~,
\label{n4}
\ee
where the integration contour in $\oint_{ob}$ runs along the obstacle boundary, encircling the obstacle from outside. The angular brackets in Eq.~(\ref{n4}) denote averaging over obstacles:
\be
\sum_i\oint_{ob}\! d{\bf l}\,(\ldots)=N\left\langle\oint_{ob}\! d{\bf l}\,(\ldots)\right\rangle~,
\label{n5}
\ee
where the sum $\sum_i\equiv\sum_{i=1}^N$ is taken over all $N$ obstacles inside the system. Note that the term in Eq.~(\ref{40a}) that is associated with the Lorentz force (namely, $-s\omega_c\psi$) drops out from Eq.~(\ref{n4}). This is because $\psi$ is constant along the boundary of an impenetrable obstacle, so that the integral $\oint_{ob}d{\bf l}\,\psi$
along the boundary
vanishes. This property of $\psi$ is similar to Eq.~(\ref{41f}) and equivalent to the condition that no current flows through the boundary, i.e., it holds independently of whether the boundary condition is diffusive or specular (we will discuss the behavior of the hydrodynamic variables around an impenetrable disk in detail in Sec.~\ref{s7}).

In Eq.~(\ref{n1}), the form of $\bf E$ inside the area need not be specified explicitly (apart from the property of $\bf E$ being curl-free); in particular, in writing Eq.~(\ref{n4}), it was sufficient to represent $V$ in terms of the hydrodynamic variables on the boundary of the liquid around the obstacles. When averaging $\bf E$ over the area fully covered by the liquid, $\bf E$ can be written in terms of the hydrodynamic variables at every point of the area integration and, after the use of Eq.~(\ref{41b}), represented as
\be
{\bf E}=-\frac{m}{e}\left[\,\nu\nabla\times\Omega{\bf e}_z+s\nabla\left(\nu_{\rm H}\Omega-\omega_c\psi\right)\,\right]~.
\label{41c}
\ee
Applying Green's theorem to $\bf E$ in the form of Eq.~(\ref{41c}) yields
\begin{align}
&\langle{\bf E}\rangle_{\rm liq}=\langle{\bf E}_{\rm H}\rangle+{\bf E}^\prime
-\frac{mN}{eS}\nonumber\\
&\times\left(\nu\left\langle\oint_{ob}\!d{\bf l}\,\Omega\right\rangle+s\nu_{\rm H}{\bf e}_z\times\left\langle\oint_{ob}\!d{\bf l}\,\Omega\right\rangle\right)~,
\label{41d}
\end{align}
where
\begin{align}
{\bf E}^\prime=\frac{m}{eS}
\left(\nu\oint_{sb}d{\bf l}\,\Omega+s\nu_{\rm H}{\bf e}_z\times\oint_{sb}d{\bf l}\,\Omega\right)
\label{41e}
\end{align}
is the viscosity-associated external-boundary term. Note that an equivalent way to state the relation between the terms proportional to $\nu$ and $\nu_{\rm H}$ in Eqs.~(\ref{41d}) and (\ref{41e}) is the identity
\be
\oint\!d{\bf l}\,\widetilde\Omega={\bf e}_z\times\oint\!d{\bf l}\,\Omega~,
\ee
valid for an arbitrary contour of integration enclosing the area (possibly multiply connected) within which $\widetilde\Omega$ is the harmonic conjugate of $\Omega$. In our problem, this means an arbitrary area fully covered by the liquid.

The contour integral of $\psi$ along the boundaries of obstacles vanishes in Eq.~(\ref{41d}), similarly to Eq.~(\ref{n4}). Note that $S$ in Eq.~(\ref{12e2}) [and, by means of Eq.~(\ref{41g}), in the Hall field $\langle{\bf E}_{\rm H}\rangle$ in Eq.~(\ref{41d})] is the total area of the system, which includes empty space inside obstacles, whereas the integration is performed over space filled with the liquid (another way to formalize the averaging is to integrate over the whole space and assign ${\bf v}=0$ to space inside voids). This is particularly important for the calculation of the Hall resistivity $\rho_{xy}$, with a deviation of $s\rho_{xy}$ from the ``universal" value $m\omega_c/e^2n$ being produced, as we discuss later in Sec.~\ref{s9}, by viscosity but not by the effect of exclusion volume.

Summing up the averaged fields inside obstacles and outside them, the total field $\langle{\bf E}\rangle$ [Eqs.~(\ref{n2}) and (\ref{n3})] is obtained as
\begin{align}
&\langle{\bf E}\rangle=\langle{\bf E}_{\rm H}\rangle+{\bf E}^\prime\nonumber\\
&-\frac{mN}{eS}\nu\left(\left\langle\oint_{ob}\!d{\bf l}\,\Omega\right\rangle+{\bf e}_z\times\left\langle\oint_{ob}\!d{\bf l}\,\widetilde\Omega\right\rangle\right)~.
\label{41h}
\end{align}
The Hall-viscosity terms in $\langle{\bf E}\rangle_{\rm obs}$ and $\langle{\bf E}\rangle_{\rm liq}$ that are proportional to ${\bf e}_z\times\langle\oint_{ob}d{\bf l}\,\Omega\rangle$ cancel out in the total field $\langle\bf E\rangle$ exactly. This cancellation emerges straightforwardly within the derivation above, but it 
signifies a nontrivial effect of Hall viscosity on the distribution of the electric field created inside and around an obstacle by the flow. We will return to this point in Sec.~\ref{s8b}.

The voltage drop on the sample is retrievable from
\be
\oint_{sb}d{\bf l}\,V=-S{\bf e}_z\times\langle{\bf E}\rangle
\label{41i}
\ee
[Eq.~(\ref{n1})], where $\langle{\bf E}\rangle$ is given by Eq.~(\ref{41h}). If the system is pushed from equilibrium by sending a current through it from an external current source, the velocity distribution with a given average $\langle{\bf v}\rangle$ [Eq.~(\ref{12e2})] generates an electric field in the bulk and a voltage on the boundary of the sample according to Eqs.~(\ref{41h}) and (\ref{41i}). Vice versa, if $\langle{\bf E}\rangle$ is viewed as a source of nonequilibrium, the friction force counterbalances the applied force according to Eq.~(\ref{41h}).

Equation (\ref{41h}) elucidates the rationale behind the representation of the first term in Eq.~(\ref{41c}) through $\Omega$ instead of $\widetilde\Omega$ [by using Eq.~(\ref{41b})]. Specifically, this made it possible to explicitly express $\langle{\bf E}\rangle$ in Eq.~(\ref{41h}) in terms of the contributions of individual obstacles, as opposed to Eq.~(\ref{41j}). The two terms in the round brackets in Eq.~(\ref{41h}), induced by dissipative viscosity, are the contributions to $\langle{\bf E}\rangle$ of the electric field in the liquid and inside the obstacles, respectively. They do not cancel in $\langle{\bf E}\rangle$, in contrast to the opposite-signed, as already mentioned above, terms associated with Hall viscosity.

It is worth emphasizing that Eqs.~(\ref{n4}), (\ref{41d}) and (\ref{41h}) are quite general, being valid in an arbitrarily shaped sample of an arbitrary size, with an arbitrary number of obstacles inside it. The shape of obstacles can also be arbitrary. This enables us to obtain the expression for $\langle{\bf E}\rangle$ in the thermodynamic limit directly from Eq.~(\ref{41h}).

\subsection{Thermodynamic limit}
\label{s6aB}

In the limit of $S\to\infty$ with the density of randomly distributed obstacles $n_d=N/S$ and the shape (``aspect ratio") of the system both held fixed---which is a definition of the thermodynamic limit in our problem---the average $\langle{\bf E}\rangle$ is fully determined, apart from $\langle{\bf E}_{\rm H}\rangle$, by the average contribution of an individual obstacle. The field ${\bf E}^\prime$, given by the contour integral $\oint_{sb}d{\bf l}\,\Omega$ along the sample boundary, vanishes in this limit, since $\Omega$ inside the liquid is bounded in this limit from above (for a stationary distribution of $\bf v$ with a given $\langle{\bf v}\rangle$). This is in contrast to the contour integral $\oint_{sb}d{\bf l}\,\widetilde\Omega$ in Eq.~(\ref{41j}), which is an extensive quantity for nonzero $n_d$. In the thermodynamic limit, Eq.~(\ref{41h}) thus becomes
\begin{align}
&\langle{\bf E}\rangle\stackrel{\begin{matrix}
{\scriptstyle\rm therm.} \\[-0.5em] {\scriptstyle\rm limit} \end{matrix}}{\longrightarrow}\langle{\bf E}_{\rm H}\rangle\nonumber\\
&-\frac{m\nu}{e}n_d\left(\left\langle\oint_{ob}\!d{\bf l}\,\Omega\right\rangle+{\bf e}_z\times\left\langle\oint_{ob}\!d{\bf l}\,\widetilde\Omega\right\rangle\right)~.
\label{41k}
\end{align}

Note that solely the dissipative viscosity coefficient $\nu$ is explicitly present in Eq.~(\ref{41k}), whereas Hall viscosity may only affect $\langle{\bf E}\rangle$ in the thermodynamic limit through the Hall-viscosity-dependent boundary conditions [Eqs.~(\ref{5}) and (\ref{6})], which then modify, for given $\langle{\bf v}\rangle$, the integrals $\langle\oint_{ob}d{\bf l}\,\Omega\rangle$ and $\langle\oint_{ob}d{\bf l}\,\widetilde\Omega\rangle$ in Eq.~(\ref{41k}). It is also worth noting that the absence of an explicit dependence of $\langle{\bf E}\rangle$ on $\nu_{\rm H}$ is a peculiarity of the thermodynamic limit, since the field ${\bf E}^\prime$, associated with the sample boundary, preserves the explicit dependence on $\nu_{\rm H}$ [Eq.~(\ref{41e})]. In what follows, $\langle{\bf E}\rangle$ is understood as calculated in the thermodynamic limit.

Equation (\ref{41k}) is a general formula for $\langle{\bf E}\rangle$ in the thermodynamic limit, written in terms of the integrals $\langle\oint_{ob}d{\bf l}\,\Omega\rangle$ and $\langle\oint_{ob}d{\bf l}\,\widetilde\Omega\rangle$ along the boundaries of arbitrarily shaped obstacles. The property of $\widetilde\Omega$ being the harmonic conjugate of $\Omega$ implies that the integral of $\widetilde\Omega$ is expressible in terms of $\Omega$ alone, and vice versa. This connection substantially simplifies for our choice of obstacles in the form of a disk (Sec.~\ref{s5}). In this case, $\langle\oint_{ob}d{\bf l}\,\widetilde\Omega\rangle$ can be represented in terms of $\Omega$ [by directly integrating by parts and using Eq.~(\ref{41b}) afterwards] as
\be
\oint_R\!d{\bf l}\,\widetilde\Omega={\bf e}_z\times\oint_R\!d{\bf l}\,({\bf r}\nabla\Omega)~,
\label{n6}
\ee
with the integration contour in $\oint_R$ running along the circumference of a circle of radius $R$. The viscosity-induced component of $\langle{\bf E}\rangle$ is then written only in terms of the vorticity as follows:
\be
\langle{\bf E}\rangle=\langle{\bf E}_{\rm H}\rangle-\frac{m\nu}{e}n_d\left\langle\oint_R\!d{\bf l}\,\left(\Omega-{\bf r}\nabla\Omega\right)\right\rangle~.
\label{41l}
\ee

Note that Eqs.~(\ref{41k}) and (\ref{41l}), describing the thermodynamic limit, do not assume that the electric field is macroscopically homogeneous. This is especially important because the distribution of the electric field in a 2D sample is generically inhomogeneous on the scale of the system size \cite{rendell81,macdonald83,thouless85}, even in the thermodynamic limit and even in the limit of ideal screening. In particular, if the sample is of a rectangular form, the distribution of the electric field depends in an essential way on the aspect ratio of the sample. Therefore, the area averaging in a given configuration of disorder, which is the strict meaning of the angular brackets in all of the above equations, reduces to local disorder averaging in the thermodynamic limit, but generically with the locally averaged integrals $\oint_{ob}$ being dependent on the obstacle position on the scale of the system size.

One example of a macroscopically homogeneous 2D flow, which we particularly have in mind, is that of an ideal (infinitely long) Hall bar, sufficiently wide to also exclude the effect of friction on the edges (present if these are not specularly reflecting). In the limit of ideal screening (zero screening length), the Hall field is then macroscopically homogeneous across the Hall bar, with the homogeneous component created by the charge density that varies on the scale of the bar width, diverging at the edges (as the inverse square root of the distance to the edge) \cite{macdonald83,thouless85,shik93}. For a long Hall bar, a weak nonideality of screening produces, apart from boundary effects (cutting off the divergency of the charge density), only a small inhomogeneity of the field in the bulk \cite{shik93}.

\subsection{Resistivity vs.\ vorticity}
\label{s6aC}

From Eq.~(\ref{41k}), the resistivity tensor $\hat\rho$, defined by the relation \cite{screened}
\be
\langle{\bf j}\rangle=\hat\rho^{-1}\langle{\bf E}\rangle
\label{6a}
\ee
between the averaged electric field $\langle\bf E\rangle$ and the averaged current density $\langle{\bf j}\rangle=-e\langle{n\bf v}\rangle$, is fully determined in the linear response limit by two vectors, $\langle{\bf v}\rangle$ and
\be
{\bf p}=\frac{m\nu}{2\pi e}\left(\left\langle\oint_{ob}\!d{\bf l}\,\Omega\right\rangle+{\bf e}_z\times\left\langle\oint_{ob}\!d{\bf l}\,\widetilde\Omega\right\rangle\right)~,
\label{22a}
\ee
both expressible in terms of the distribution of the hydrodynamic velocity,
with
\be
\langle{\bf E}\rangle-\langle{\bf E}_{\rm H}\rangle=-2\pi n_d{\bf p}~.
\label{22x}
\ee
Specifically, a macroscopically isotropic system is characterized by
\begin{align}
&\rho_{xx}=\frac{2\pi}{e}\frac{n_d}{n_0}\,\frac{{\bf p}\langle{\bf v}\rangle}{\langle{\bf v}\rangle^2}~,
\label{21}\\
&s\rho_{xy}-\frac{m\omega_c}{e^2n_0}=-\frac{2\pi}{e}\frac{n_d}{n_0}\,\frac{\left({\bf p}\times{\bf n}\right)\langle{\bf v}\rangle}{\langle{\bf v}\rangle^2}~.
\label{22}
\end{align}
This is an exact (linear response) expression for $\hat\rho$ in hydrodynamics for the case of a homogeneous equilibrium density $n_0$ between obstacles. In this expression, $\hat\rho$ is fully---apart from the universal term in the left-hand side of Eq.~(\ref{22})---determined by the vorticity and its harmonic conjugate on the boundary of obstacles. For circular obstacles, $\bf p$ can also be represented as [Eq.~(\ref{n6})]
\be
{\bf p}=\frac{m\nu}{2\pi e}\left\langle\oint_R\!d{\bf l}\,\left(\Omega-{\bf r}\nabla\Omega\right)\right\rangle~.
\label{22b}
\ee
As follows from Eq.~(\ref{21}), the projection of $\bf p$ on $\langle{\bf v}\rangle$ in a stable flow is positive for arbitrary relation between $\nu$ and $\nu_{\rm H}$.

If there are random macroscopic inhomogeneities of the obstacle density (or any other characteristics of disorder for that matter), Eqs.~(\ref{21}) and (\ref{22}) are still valid with $n_d$ understood as the average density in the thermodynamic limit and the vector $\bf p$ averaged according to Eq.~(\ref{n5}). It is perhaps also worth mentioning that $\hat\rho^{-1}$ in Eq.~(\ref{6a}) should then not be confused with the local conductivity averaged over the inhomogeneities \cite{lanlif8,herring60,stroud76}.

\section{Single obstacle}
\label{s7}

The field $\bf v$, constrained by Eqs.~(\ref{4}) or (\ref{5}), is infrared-singular in the 2D case \cite{lamb11,lanlif6}. For a single obstacle, the logarithmic hydrodynamic singularity can be regularized by going beyond the linear response theory (by retaining the inertia term in the Navier-Stokes equation; in particular, by means of the conventional Oseen approximation \cite{lamb11,proudman57}). Another possibility is to introduce a total-momentum relaxation by additional weak disorder \cite{lucas17}. These regularizators are not necessary if there is a finite density of obstacles, as in our model (or in the model with pointlike scatterers from Refs.~\cite{hruska02,guo16}), where a characteristic spatial scale $L_v$ (we will discuss it in Sec.~\ref{s9}) emerges on which the viscous effect produced by a given obstacle fades away because of the presence of other obstacles. In a dilute obstacle array, with $n_dR^2\ll 1$, the thus defined relaxation length $L_v$ is much larger than $R$.

With this rationale in mind, our strategy for finding the average solution to Eqs.~(\ref{2}) and (\ref{3}) is, then, to first solve an auxiliary problem of the flow past a single void by imposing an inhomogeneous boundary condition on the circle of radius $L$ around it. Let us first choose
\be
{\bf v}({\bf r})\vert_{r=L}={\bf v}_0
\label{9}
\ee
with an arbitrary velocity ${\bf v}_0$, constant along the circle (centered at the origin). The solution to Eqs.~(\ref{2}) and (\ref{3}) is to be found within the space specified by $R<r<L$, with one of two conditions fixed at $r=R$ [Eqs.~(\ref{4}) or (\ref{5})] and the other at $r=L$ [Eq.~(\ref{9})].

For our choice of the boundary conditions, the first angular harmonic of $\psi({\bf r})$ decouples \cite{bc_harmonics}, so that the Navier-Stokes equation reduces to an ordinary differential equation for the complex function $\chi(r)$, dependent on the radial coordinate only and defined by
\be
\psi({\bf r})={\rm Re}\left\{\chi(r)e^{i\varphi}\right\}~.
\label{12}
\ee
Specifically, from Eqs.~(\ref{11}) and (\ref{12}), $\chi$ obeys
\be
\left[\frac{d}{dr}\left(\frac{d}{dr}+\frac{1}{r}\right)\right]^2\chi=0~.
\label{13}
\ee

The radial and tangential components of ${\bf v}$ and the radial-tangential component of the momentum flux tensor are given, in terms of $\chi$, by
\begin{align}
v_r&=-\frac{1}{r}\,{\rm Im}\left\{\chi e^{i\varphi}\right\}~,\quad v_\varphi=-{\rm Re}\left\{\chi^\prime e^{i\varphi}\right\}~,\label{14}\\
\Pi_{r\varphi}&={\rm Re}\left\{ \left[\,\nu\chi^{\prime\prime}-(\nu+2is\nu_{\rm H})\left(\frac{\chi}{r}\right)^\prime\,\right]e^{i\varphi}\right\}~,
\label{15a}
\end{align}
where the sign $'$ denotes the derivative $d/dr$. The counterparts of the boundary conditions (\ref{4}) and (\ref{5}) are, then, rewritten in terms of $\chi$ as
\be
\chi(R)=0~,\quad\!\chi^\prime(R)=0
\label{16}
\ee
for the former case and
\be
\chi(R)=0~,\quad\!R\chi^{\prime\prime}(R)-\left(1+2is\,\frac{\nu_{\rm H}}{\nu}\right)\!\chi^\prime(R)=0
\label{17a}
\ee
for the latter. On the outer boundary, Eq.~(\ref{9}) translates into
\be
\chi(L)=-iL|{\bf v}_0|e^{-i\varphi_0}~,\quad \left(\chi/r\right)^\prime_{r=L}=0~,
\label{18a}
\ee
where $\varphi_0$ is the polar angle of ${\bf v}_0$ [note that the second equation (\ref{18a}) means, because of Eq.~(\ref{14}), the vanishing of $v_r^\prime$ at $r=L$]. The two conditions on the 2D vector ${\bf v}({\bf r})$ defined in the 2D plane, fixed on the inner and outer boundaries [Eqs.~(\ref{4}) or (\ref{5}), and Eq.~(\ref{9})], along with the constraint (\ref{2}) impose altogether four conditions [Eqs.~(\ref{16}) or (\ref{17a}), and Eqs.~(\ref{18a})] on the complex scalar $\chi(r)$.

The general solution to Eq.~(\ref{13}) is a sum of four terms
\be
\chi=R\left(C_1\bar{r}\ln \bar{r}+C_2\bar{r}^3+C_3\bar{r}+C_4/\bar{r}\right)~,
\label{19b}
\ee
where $\bar{r}=r/R$ and the complex numbers $C_{1,2,3,4}$ are to be fixed by the boundary conditions on the inner and outer circles. For $\psi$ from Eq.~(\ref{12}), it is convenient to represent $\bf v$ as a sum of the zeroth and second angular harmonics
\be
{\bf v}=\left(
\begin{array}{c}
{\rm Re}\\
{\rm Im}\\
\end{array}
\right)\!\left\{g_+^*(r)+g_-(r)e^{2i\varphi}\right\}~,
\label{26}
\ee
(here and below, the upper and lower lines correspond to the $x$ and $y$ vector components, respectively), where $g_\pm$ are given by the combinations
\be
g_\pm=\frac{i}{2}\left(\pm\chi^\prime+\frac{\chi}{r}\right)
\label{27a}
\ee
(related by $g_+'+g_-'+2g_-/r=0$). Note that the terms with $C_3$ and $C_4$ drop out from $\Omega$ and $\nabla^2{\bf v}$ (and $g_+'$, which fully determines $\Omega$ and $\nabla^2{\bf v}$, for that matter). That is, the vorticity and the viscous force around the obstacle depend on only two constants, $C_1$ and $C_2$. Specifically,
\begin{align}
&\Omega=-\frac{2}{R}\,{\rm Re}\left\{\left(\frac{C_1}{\bar{r}}+4C_2\bar{r}\right)e^{i\varphi}\right\}~,
\label{24a}\\
&\nabla^2\!{\bf v}=\frac{2}{R^2}\!\left(
\begin{array}{c}
{\rm Im}\\
{\rm Re}\\
\end{array}
\right)\!
\left\{\frac{1}{\bar{r}^2}C_1^*e^{-2i\varphi}-4C_2\right\}~.
\label{24b}
\end{align}
The contour integral of $\Omega$ along the boundary of the obstacle reads
\begin{align}
\oint_R\!d{\bf l}\,\Omega=-2\pi
\!\left(
\begin{array}{c}
{\rm Im}\\
{\rm Re}\\
\end{array}
\right)\!
\left\{C_1+4C_2\right\}~.
\label{25}
\end{align}

As seen from the singular terms in Eqs.~(\ref{24a}) and (\ref{24b}), the constant $C_1$ determines the strength of Stokes' singularity. The constant $C_2$ gives the average of the viscous force over the space filled with the liquid. Indeed, $C_1$ drops out from the integral $\int\!d^2{\bf r}\,\nabla^2{\bf v}$, with $\nabla^2{\bf v}$ from Eq.~(\ref{24b}), over the area with $R<r<L$. This can also be seen by noting that this area integral is given by the difference of $\oint\! d{\bf l}\,\Omega$ taken on the outer ($r=L$) and inner ($r=R$) circles.

As follows from a comparison of Eqs.~(\ref{16}) and (\ref{17a}), the difference between the specular and diffusive boundary conditions disappears for $\nu_{\rm H}/\nu\to\infty$:
\be
\frac{\nu_{\rm H}}{\nu}\to\infty \quad\mapsto\quad {\rm (\,specular\,\, b.c.\,\to\, diffusive\,\, b.c.\,)}
\label{25a}
\ee
This can also be inferred from Eqs.~(\ref{4})-(\ref{6}), where the only difference that remains in this limit between the specular and diffusive boundary conditions is that $v_\varphi=0$ in the latter case, whereas $v_\varphi$ can be an arbitrary constant along the boundary of an obstacle immersed in an incompressible liquid in the former. For the boundary conditions (\ref{16})-(\ref{18a}), this constant is zero, so that the difference is irrelevant. Below, we will present the results for the specular boundary condition and make use of Eq.~(\ref{25a}) to immediately obtain those for the diffusive one.

For $L/R\gg 1$, the constants $C_{1,2}$, which determine $\Omega$ and $\nabla^2{\bf v}$ in Eqs.~(\ref{24a}) and (\ref{24b}), and the functions $g_\pm(r)$ from Eq.~(\ref{26}) are written for the specular boundary condition as follows:
\begin{align}
&C_1\simeq -\frac{i}{\ln (L/R)-(1+h)/2}\,|{\bf v}_0|e^{-i\varphi_0}~,
\label{35a}\\
&C_2\simeq -\frac{1}{2}\left(\frac{R}{L}\right)^2C_1~,
\label{32}
\end{align}
and
\begin{align}
g_+\simeq iC_1\times
\left\{
\begin{array}{ll}
\ln\dfrac{r}{R}+\dfrac{1-h}{2}~, & r\ll L~, \\ \\
\ln\dfrac{r}{R}+\dfrac{1-h}{2}-\left(\dfrac{r}{L}\right)^2~,\,\, & r\gg R~,
\end{array}
\right.
\label{33a}
\end{align}
\begin{align}
g_-\simeq -\frac{i}{2}C_1\times
\left\{
\begin{array}{ll}
1-h(R/r)^2~,\quad & r\ll L~, \\ \\
1-(r/L)^2~, & r\gg R~
\end{array}
\right.
\label{36}
\end{align}
in the overlapping regions $r\ll L$ and $r\gg R$, where the terms proportional to
\be
h=\frac{is\nu_{\rm H}}{\nu+is\nu_{\rm H}}
\label{34a}
\ee
arise because of Hall viscosity. To avoid cumbersomeness, we omitted terms in the denominator of $C_1$ and in $g_\pm/C_1$ of the order of $(R/L)^2$ and higher, and the terms in $C_2/C_1$ of the order of $(R/L)^4$ and higher. In view of Eq.~(\ref{25a}), $C_{1,2}$ and $g_\pm(r)$ for the diffusive boundary condition are obtained by setting $h=1$. The ratio $C_2/C_1$ to order $(R/L)^2$ is the same for impenetrable obstacles irrespective of whether the boundary condition is diffusive or specular.

For the case of the specular boundary condition, retaining the nonlogarithmic terms in $g_+$ and $1/C_1$ is important in two respects. Firstly, it is because of the nonlogarithmic terms that $g_+$ in Eq.~(\ref{33a}) and $g_-$ in Eq.~(\ref{36}) do not vanish for $h\neq 1$ at $r=R$. Secondly, these terms change the phases of $g_\pm$ and $C_1$. The phase change is due to the presence of $\nu_{\rm H}$ in the boundary condition (\ref{17a}). Note that, despite $\bf v$ being finite at $r=R$ for the specular boundary condition, the separation of the first angular harmonic in $v_\varphi$ in Eq.~(\ref{14})
[cf.\ Eq.~(\ref{26})]
ensures that the circulation of $\bf v$ around the void $\oint\! d{\bf l}\,{\bf v}$ vanishes, also in the presence of the Hall viscosity term in Eq.~(\ref{17a}).

Interestingly, Hall viscosity makes the liquid slow down on the boundary of an obstacle with the specular boundary condition, leading to ${\bf v}=0$ on the boundary in the limit of a strong magnetic field [$h=1$ in Eqs.~(\ref{33a}) and (\ref{36})]. This is quite apart from the logarithmic suppression of $C_1$ (which constitutes Stokes' paradox) in Eq.~(\ref{35a}).

From Eqs.~(\ref{25}) and (\ref{35a}), neglecting $C_2$ compared to $C_1$ because of the relation (\ref{32}), the integral of $\Omega$ around the void for $\ln (L/R)\gg 1$ is written to leading order as
\be
\oint_R\! d{\bf l}\,\Omega\simeq\frac{2\pi}{\ln(L/R)}\,{\bf v}_0
\label{37}
\ee
for both the diffusive and specular boundary conditions. The difference between the two occurs at order $1/\ln^2(L/R)$, at which the factor ${\bf v}_0$ in Eq.~(\ref{37}) is modified as ${\bf v}_0\to{\bf v}_0+\Delta{\bf v}$, where
\begin{align}
\Delta{\bf v}=&\,\,\frac{1}{2\ln (L/R)}\nonumber\\
\times&\left\{{\bf v}_0+\frac{1}{\nu^2+\nu_{\rm H}^2}\left[\,\nu_{\rm H}^2{\bf v}_0+\nu\nu_{\rm H}\left({\bf v}_0\times{\bf n}\right)\,\right]\right\}
\label{41}
\end{align}
for the case of the specular boundary condition, and $\Delta {\bf v}$ for the case of the diffusive boundary condition is obtainable from Eq.~(\ref{41}) by, as discussed above, sending $\nu_{\rm H}/\nu\to\infty$. Notably, if $\nu_{\rm H}$ is finite (neither zero nor infinite), $\Delta{\bf v}$ is not parallel to ${\bf v}_0$ for the specular boundary condition.

The average of $\bf v$ over the area between the inner and outer circles, which is fully determined by $\psi$ on the outer boundary,
$\int_{r<L}\!d^2{\bf r}\,{\bf v}=-\oint_{r=L}\!d{\bf l}\,\psi$, is given, as follows from the first condition in Eq.~(\ref{18a}), by
\be
\int_{r<L}\!d^2{\bf r}\,{\bf v}=\pi L^2{\bf v}_0~,
\label{39}
\ee
as if there was no obstacle. That is, for this boundary condition at $r=L$, the slowing down of the liquid near the obstacle and the exclusion volume are exactly compensated in the average of $\bf v$ by an acceleration, with respect to ${\bf v}_0$, away from the obstacle \cite{vel_max}. Note that, in contrast to Eq.~(\ref{41}), the average (\ref{39}) is parallel to ${\bf v}_0$ also in the case of the specular boundary condition with $\nu_{\rm H}\neq 0$.

For $\Omega$ from Eq.~(\ref{24a}), $\widetilde\Omega$ [Eq.~(\ref{41b})] reads
\be
\widetilde\Omega=\frac{2}{R}\,{\rm Im}\left\{\left(\frac{C_1}{\bar{r}}-4C_2\bar{r}\right)e^{i\varphi}\right\}~.
\label{70}
\ee
Substituting Eq.~(\ref{70}) in the integral of $\widetilde\Omega$ along the boundary of the obstacle from Eq.~(\ref{22a}) [or, equivalently, using Eq.~(\ref{n6})] gives
\begin{align}
{\bf e}_z\times\oint_R\!d{\bf l}\,\widetilde\Omega=-2\pi
\!\left(
\begin{array}{c}
{\rm Im}\\
{\rm Re}\\
\end{array}
\right)\!
\left\{C_1-4C_2\right\}~,
\label{41m}
\end{align}
which is only different from Eq.~(\ref{25}) by the sign in front of $C_2$. The constant $C_2$ thus cancels out in the sum of the integrals of $\Omega$ and $\widetilde\Omega$ in Eq.~(\ref{22a}), so that the sum is fully determined by the constant $C_1$:
\be
\oint_R\!d{\bf l}\,\Omega+{\bf e}_z\times\oint_R\!d{\bf l}\,\widetilde\Omega=-4\pi
\!\left(
\begin{array}{c}
{\rm Im}\\
{\rm Re}\\
\end{array}
\right)
C_1~.
\label{71}
\ee
Since $C_2\ll C_1$ [Eq.~(\ref{32})], the two terms in Eq.~(\ref{71}) contribute almost equally to the sum.

For a circular obstacle, the integration in Eqs.~(\ref{25}) and (\ref{41m}) ``filters out" all but the first angular harmonic in $\Omega$ on the boundary of the obstacle. Consequently, all but the zeroth and second harmonics of $v_{x,y}$ on the outer boundary at $r=L$ are decoupled from $\oint_R\!d{\bf l}\,\Omega$ and $\oint_R\!d{\bf l}\,({\bf r}\nabla\Omega)$ [Eq.~(\ref{n6})].
Recall that only the zeroth harmonic of $v_{x,y}$ is present in the boundary condition specified by Eq.~(\ref{9}). Let us now relax this condition by adding a nonzero second harmonic of $v_{x,y}$, i.e., by fixing both $g_\pm(L)\neq 0$ [Eq.~(\ref{26})]. For
\be
g_+(L)=|{\bf v}_0|e^{-i\varphi_0}~,
\label{57b}
\ee
as in Eq.~(\ref{9}), and
\be
g_-(L)=|{\bf v}_2|e^{-i\varphi_2}~,
\label{57a}
\ee
where $\varphi_2$ is the angle of ${\bf v}_2$, Eq.~(\ref{35a}) is modified by the addition of $g_-(L)$ as
\be
C_1\simeq -i\frac{g_+(L)+2g_-(L)}{\ln(L/R)-(1+h)/2}~.
\label{56}
\ee
The accuracy of Eq.~(\ref{56}) is the same as for Eq.~(\ref{35a}) [terms of the order of $(R/L)^2$ and higher in the denominator are neglected].

For given $g_\pm(L)$, Eq.~(\ref{32}) for the relation between $C_1$ and $C_2$ is modified, to first order in $(R/L)^2$ for $C_2$ (up to the logarithmic factor in $C_1$), as
\be
C_2\simeq -\frac{1}{2}\left(\frac{R}{L}\right)^2[\,C_1-2ig_-(L)\,]~.
\label{56a}
\ee
The contribution of $C_2$ to $\oint_R\!d{\bf l}\,\Omega$ and $\oint_R\!d{\bf l}\,({\bf r}\nabla\Omega)$ is seen to remain negligibly small for $L/R\gg 1$. Equations (\ref{37}) and (\ref{41}) for $\oint_R\!d{\bf l}\,\Omega$ are modified by the shift ${\bf v}_0\to{\bf v}_0+2{\bf v}_2$. Similarly, the average (\ref{39}) of ${\bf v}({\bf r})$ over the area with $r<L$ is modified by shifting ${\bf v}_0\to{\bf v}_0+{\bf v}_2$.

\section{Resistivity tensor and Hall viscosity}
\label{s9}

Having established the general framework for the calculation of the resistance in hydrodynamics with impenetrable obstacles (Sec.~\ref{s6a}) and discussed in detail the flow around a single circular obstacle (Sec.~\ref{s7}), we proceed to calculate the resistivity tensor $\hat\rho$ of a liquid flowing through a random array of obstacles. According to Eqs.~(\ref{21}) and (\ref{22}), $\hat\rho$ is determined by two averages: the vector $\bf p$, which is an average over obstacles [Eqs.~(\ref{22a}) and (\ref{22b})], and the space averaged velocity $\langle{\bf v}\rangle$. It is convenient to view ${\bf p}\neq 0$ as a linear response, which is to be found, to the perturbation of equilibrium parametrized by a given $\langle{\bf v}\rangle$. We now use the results obtained for $\oint_R\! d{\bf l}\,\Omega$ and $\oint_R\! d{\bf l}\,\widetilde\Omega$ within the single-obstacle ``scattering problem," Eqs.~(\ref{25}) and (\ref{41m}).

\subsection{Disorder averaging}
\label{s9a}

As mentioned already at the beginning of Sec.~\ref{s7}, the rationale behind using the results of the single-obstacle model, which has an outer boundary, is based on the large separation of the spatial scales $L_v$ (the precise expression for which is to be obtained below) and $R$ in a dilute array of obstacles ($n_dR^2\ll 1$). For the flow past an individual obstacle, we therefore accurately treat the flow on spatial scales around the obstacle that are much smaller than $L_v$ and treat the rest of the system essentially as an effective medium which regularizes the singularity on the scale of $L_v$. This amounts to a ``logarithmically accurate" theory. For our problem with $\nu_{\rm H}\neq 0$, the logarithmic accuracy for the case of specular boundary conditions means that any term in $\hat\rho$ linear in $n_d$ (up to logarithmic factors dependent on $n_d$) that is distinct in powers of $\nu$ and $\nu_{\rm H}$ can be found exactly to leading order in powers of $1/{\cal L}$, where
\be
{\cal L}=2\ln(L_v/R)~.
\label{9a}
\ee

Importantly, the leading order here does not necessarily imply the first power of $1/{\cal L}$. As we will see below, the terms that do not depend on $\nu_{\rm H}$ are accurately obtainable within this approach only to order ${\cal O}(1/{\cal L})$, whereas those induced by Hall viscosity to order ${\cal O}(1/{\cal L}^2)$. The latter is the leading order at which the terms dependent on $\nu_{\rm H}$ emerge in the expansion of $\hat\rho$ in powers of $1/{\cal L}$.

Specifically, the disorder averaging in Eq.~(\ref{22a}) implies the summation (\ref{n5}) of the contributions of different obstacles, each of which can be viewed as given by Eq.~(\ref{71}) with fluctuating coefficients $C_1$:
\begin{align}
{\bf p}=-\frac{2m\nu}{e}
\!\left(
\begin{array}{c}
{\rm Im}\\
{\rm Re}\\
\end{array}
\right)\!
\langle C_1\rangle~.
\label{57}
\end{align}
One can visualize the averaging of $C_1$ by placing each obstacle, in a given realization of disorder, at the center of a circle of radius $L_v$.  The argument that Eq.~(\ref{56}) for a single obstacle with $L\sim L_v$ can be used to accurately calculate the leading contributions to $\bf p$---also induced by Hall viscosity---proceeds by considering the characteristic orders of magnitude of $f_\pm(L_v)$ on such a circle with the use of Eqs.~(\ref{33a}) and (\ref{36}). The relevant orders of magnitude in powers of $1/{\cal L}$ are as follows.

Firstly, the main contribution to $g_+(L_v)$ in a random environment is given by Eq.~(\ref{57b}) with ${\bf v}_0=\langle{\bf v}\rangle$, up to corrections that emerge at order ${\cal O}(1/{\cal L})$ but on average do not rotate ${\bf v}_0$ with respect to $\langle{\bf v}\rangle$ at this order. Secondly, the leading contribution to $g_-(L_v)$ is of the order of $g_+(L_v)/{\cal L}$, and the phases of $g_\pm(L_v)$ are on average the same at this order, i.e., both the average ${\bf v}_0$ and ${\bf v}_2\sim {\bf v}_0/{\cal L}$, with ${\bf v}_2$ from Eq.~(\ref{57a}), are parallel to $\langle{\bf v}\rangle$ to order ${\cal O}(1/{\cal L})$. Thirdly, both the amplitudes and the phases of $g_\pm(L_v)$ are affected by Hall viscosity, with the average ${\bf v}_0$ and ${\bf v}_2$ rotated with respect to $\langle{\bf v}\rangle$, only at order ${\cal O}(1/{\cal L}^2)$.

An important aspect of this argument is that the disorder averaging preserves the combination $h$ [Eq.~(\ref{34a})] in which the ratio $\nu_{\rm H}/\nu$ appears in $C_1$ in Eq.~(\ref{56}) and, therefore, in $\Delta{\bf v}$ in Eq.~(\ref{41}). With the orders of magnitude obtained above, the key point with regard to the dependence of $\langle C_1\rangle$ on $\nu_{\rm H}$ is that this dependence emerges at order ${\cal O}(1/{\cal L}^2)$ and, to this order, it comes from the denominator of Eq.~(\ref{56}) but not from the dependence of $g_\pm(L_v)$ on $\nu_H$. Moreover, to order ${\cal O}(1/{\cal L}^2)$, only $g_+(L_v)$ contributes to the amplitude of the $\nu_H$ dependent term in $\langle C_1\rangle$ but not $g_-(L_v)$.

With this input, Eq.~(\ref{56}) produces
\be
\langle C_1\rangle=\langle C_1\rangle_0+\langle C_1\rangle_{\rm H}~,
\label{60b}
\ee
where $\langle C_1\rangle_0$ does not depend on $\nu_{\rm H}$ and $\langle C_1\rangle_{\rm H}$ is induced by Hall viscosity, with the leading (in powers of $1/{\cal L}$) contributions to $\langle C_1\rangle_{\rm 0,H}$ given by:
\begin{align}
&\langle C_1\rangle_0\simeq -\frac{2i}{\cal L}\,|\langle{\bf v}\rangle|e^{-i\varphi_{\langle{\bf v}\rangle}}~,
\label{58}\\
&\langle C_1\rangle_{\rm H}\simeq -\frac{2ih}{{\cal L}^2}\,|\langle{\bf v}\rangle|e^{-i\varphi_{\langle{\bf v}\rangle}}~.
\label{59}
\end{align}
The angle $\varphi_{\langle{\bf v}\rangle}$ specifies the direction of $\langle{\bf v}\rangle$. Representing the vector $\bf p$ [Eq.~(\ref{57})] as a sum of two components,
\be
{\bf p}={\bf p}_0+{\bf p}_{\rm H}~,
\label{60a}
\ee
similarly to $\langle C_1\rangle$, Eqs.~(\ref{58}) and (\ref{59}) correspond to
\begin{align}
&{\bf p}_0\simeq \frac{4m\nu}{e{\cal L}}\,\langle{\bf v}\rangle~,
\label{60}\\
&{\bf p}_{\rm H}\simeq \frac{4m\nu}{e{\cal L}^2}\,\frac{1}{\nu^2+\nu_{\rm H}^2}\left[\,\nu_{\rm H}^2\langle{\bf v}\rangle+\nu\nu_{\rm H}\left(\langle{\bf v}\rangle\times{\bf n}\right)\,\right]~.
\label{61}
\end{align}
Note that, in accordance with the above argument, ${\bf p}_{\rm H}$ is precisely given by the $\nu_{\rm H}$ dependent component of $\Delta{\bf v}$ [Eq.~(\ref{41})] with ${\bf v}_0=\langle{\bf v}\rangle$. The two terms in ${\bf p}_{\rm H}$, one parallel to $\langle{\bf v}\rangle$ and the other perpendicular to it, are thus inherently related to each other.

It is notable that, as follows from Eqs.~(\ref{25}), (\ref{41m}), (\ref{71}), and specifically the comment right below Eq.~(\ref{71}), the contributions to $\bf p$ of the interior space of obstacles and the space outside them are equal to each other in the limit of a dilute array. This is because of the relation (\ref{56a}) between the characteristic values of $C_2$ and $C_1$ for $L$ substituted by the average distance between obstacles $n_d^{-1/2}$, or the still undetermined relaxation length $L_v$ for that matter. Correspondingly,
\be
\langle{\bf E}\rangle_{\rm obs}{\bf p}\simeq\langle{\bf E}\rangle_{\rm liq}{\bf p}\simeq\langle{\bf E}\rangle{\bf p}/2~,
\label{61a}
\ee
where the sign $\simeq$ becomes ``equals" at linear order in $n_d$ (and arbitrary order in $1/{\cal L}$).

\subsection{Relaxation length}
\label{s9b}

It remains to find $L_v$. Recall that $L_v$, which is the infrared cutoff in the logarithm ${\cal L}$ [Eq.~(\ref{9a})], is defined as the relaxation length for the viscous effect produced by a given obstacle in the random environment of other obstacles. One way to obtain an estimate for $L_v$ is to define it by representing the average friction force, i.e., the component of $e\langle{\bf E}\rangle$ that is parallel to $\langle{\bf v}\rangle$, as $-(m\nu/L_v^2)\langle{\bf v}\rangle$:
\be
e\frac{\langle{\bf E}\rangle\langle{\bf v}\rangle}{\langle{\bf v}\rangle^2}=-\frac{m\nu}{L_v^2}~,
\label{64}
\ee
which means the damping rate
\be
1/\tau=\nu/L_v^2
\label{64a}
\ee
for the current. On the other hand, as follows from Eqs.~(\ref{41k}), (\ref{71}), and (\ref{58}), to order ${\cal O}(1/{\cal L})$, the amplitude of the friction force obeys
\be
e\frac{\langle{\bf E}\rangle\langle{\bf v}\rangle}{\langle{\bf v}\rangle^2}\simeq -8\pi \frac{mn_d\nu}{\cal L}~.
\label{65}
\ee
By combining Eqs.~(\ref{64}) and (\ref{65}), the equation to find $L_v$ takes the form $1/L_v^2\simeq 8\pi n_d/{\cal L}$, which yields, with logarithmic accuracy,
\be
L_v\simeq\sqrt{\frac{\ln(1/n_dR^2)}{8\pi n_d}}~.
\label{62}
\ee

In a dilute obstacle array, $L_v$ is seen to be much larger [by the logarithmic factor in Eq.~(\ref{62})] than the average distance between obstacles. That is, it takes as many as $\ln(1/n_dR^2)$ obstacles to ``blur" the viscous effect of one obstacle. With logarithmic accuracy, though, $\cal L$ can be approximated as $\ln(1/n_dR^2)$ [by neglecting the double-logarithm term $\ln\ln(1/n_dR^2)$ in ${\cal L}$].

\subsection{Resistivity tensor}
\label{s9c}

Having found $\bf p$ [Eqs.~(\ref{60a})-(\ref{61})] and specified $L_v$ [Eq.~(\ref{62})] in the definition of $\cal L$, we can now use Eqs.~(\ref{21}) and (\ref{22}) to obtain the resistivity tensor $\hat\rho$. It is convenient to split the dissipative resistivity $\rho_{xx}$ into two parts as
\be
\rho_{xx}=\rho_{xx,0}+\rho_{xx,{\rm H}}~,
\label{66}
\ee
where $\rho_{xx,0}$ is independent of $\nu_{\rm H}$ and $\rho_{xx,{\rm H}}$ emerges entirely because of Hall viscosity, in line with the corresponding representation of $\langle C_1\rangle$ [Eq.~(\ref{60b})] and $\bf p$ [Eq.~(\ref{60a})]. For the specular boundary condition, the leading terms in $\rho_{xx,0}$ and $\rho_{xx,{\rm H}}$ are then given by
\begin{align}
&\rho_{xx,0}\simeq \frac{8\pi m}{e^2{\cal L}}\,\frac{n_d}{n_0}\,\nu~,
\label{67a}\\
&\rho_{xx,{\rm H}}\simeq \frac{8\pi m}{e^2{\cal L}^2}\,\frac{n_d}{n_0}\,\frac{\nu\nu_{\rm H}^2}{\nu^2+\nu_{\rm H}^2}~.
\label{67b}
\end{align}
The leading term in the deviation of the Hall resistivity $\rho_{xy}$ from the universal value, for the specular boundary condition, is written as
\begin{align}
s\rho_{xy}-\frac{m\omega_c}{e^2n_0}\simeq \frac{8\pi m}{e^2{\cal L}^2}\,\frac{n_d}{n_0}\,\frac{\nu^2\nu_{\rm H}}{\nu^2+\nu_{\rm H}^2}~.
\label{67c}
\end{align}
It may be worth noting that $\rho_{xx}$ in hydrodynamics with obstacles is finite for the no-stress condition on their boundaries, in contrast to the Poiseuille-like flow through a straight pipe with this boundary condition.

Equations~(\ref{67a})-(\ref{67c}) were obtained by relying on the exact hydrodynamic formula we derived for $\hat\rho$ (Sec.~\ref{s6a}), the exact solution of the single-obstacle problem (Sec.~\ref{s7}), and the use of the large parameter ${\cal L}\gg 1$ to perform disorder averaging (in the preceding part of Sec.~\ref{s9}). This allowed us to proceed by making precisely controlled approximations [to order ${\cal O}(1/{\cal L})$ in Eq.~(\ref{67a}) and to order ${\cal O}(1/{\cal L}^2)$ in Eqs.~(\ref{67b}) and (\ref{67c})] but required a somewhat complex logical construction with regard to the disorder averaging below Eq.~(\ref{57}). In Appendix~\ref{a}, we complement the foregoing calculation by using the mean-field approximation to describe the effect produced by the random environment of a given obstacle on the flow around it. This approximation, while giving an explicit construction for calculating the disorder-averaged flow around a given obstacle, reproduces Eqs.~(\ref{67a})-(\ref{67c}) and has exactly the same range of applicability with respect to the accuracy of the expansion of $\hat\rho$ in powers of $1/{\cal L}$ as the above calculation.

Let us now discuss the significance of Eqs.~(\ref{67a})-(\ref{67c}). To begin with, recall that $\hat\rho$ for the diffusive boundary condition is obtainable from $\hat\rho$ for the specular boundary condition as the limit $\nu_{\rm H}/\nu\to\infty$ [Eq.~(\ref{25a})]. Note that the leading term (\ref{67a}) in the expansion of $\rho_{xx}$ in powers of $1/{\cal L}$ does not depend on $\nu_{\rm H}$ and thus is the same for an arbitrary degree of specularity in the boundary condition. Importantly, however, as mentioned already at the end of Sec.~\ref{s4} and detailed in Sec.~\ref{s7}, Hall viscosity modifies the flow ${\bf v}({\bf r})$ in the case of the specular boundary condition. Remarkably, as follows from Eq.~(\ref{67b}), the change of the flow by Hall viscosity---with Hall viscosity being by itself  dissipationless in the bulk of the flow---affects the dissipative resistivity. This occurs at order ${\cal O}(1/{\cal L}^2)$, with $\rho_{xx,{\rm H}}$ relying on the interplay of Hall and dissipative viscosity, thus vanishing for both $\nu_{\rm H}=0$ for arbitrary $\nu$ and $\nu=0$ for arbitrary $\nu_{\rm H}$.

Since, as discussed in Sec.~\ref{s4}, neither the Lorentz force nor the Hall viscosity force in the bulk affect the flow, the vector $\bf p$ in Eq.~(\ref{22a}) is only rotated by Hall viscosity with respect to $\langle{\bf v}\rangle$---and thus produces a finite contribution to $\rho_{xy}$---if the boundary condition depends on $\nu_{\rm H}$. In accordance with this, the deviation of $\rho_{xy}$ in Eq.~(\ref{67c}) from the universal value vanishes in the limit $\nu_{\rm H}/\nu\to\infty$ [Eq.~(\ref{25a})]. Otherwise, the nonuniversal term in $\rho_{xy}$ relies on the interplay of Hall and dissipative viscosity, being zero for $\nu_{\rm H}=0$ and $\nu\neq 0$ and for $\nu=0$ and  $\nu_{\rm H}\neq 0$, similarly to $\rho_{xx,{\rm H}}$.

\subsection{Hall coefficient}
\label{s9d}

The structure of the tensor $\hat\rho$ in Eqs.~(\ref{21}) and (\ref{22}) suggests that Hall viscosity can be thought of as leading to the emergence of an effective magnetic field in the liquid, linear in $n_d$, which amounts to ``screening" of the external magnetic field and, to order ${\cal O}(1/{\cal L}^2)$ in the effective field, results in a shift $\omega_c\to\omega_c+\Delta\omega_c$ in $\hat\rho$, where
\be
\Delta\omega_c=\frac{8\pi n_d}{{\cal L}^2}\,\frac{\nu^2\nu_{\rm H}}{\nu^2+\nu_{\rm H}^2}~.
\label{54}
\ee
Since $\Delta\omega_c/\omega_c>0$, this effect is rather that of overscreening. Note that $\Delta\omega_c$ depends on the electron density $n_0$ only through the dependence of the viscosity tensor on it.

The Hall coefficient (``Hall constant") $R_{\rm H}=-s\rho_{xy}/B$, where $B$ is the modulus of the amplitude of the magnetic field, is thus modified by Hall viscosity. It is worth emphasizing that $B$ in the above definition of $R_{\rm H}$ is the external (not effective) field. To order ${\cal O}(1/{\cal L}^2)$, we have from Eq.~(\ref{54}):
\be
R_{\rm H}\simeq -\frac{1}{en_0c}\left(1+\frac{\Delta\omega_c}{\omega_c}\right)~,
\label{50}
\ee
where $c$ is the speed of light. In the small-$B$ limit, using $\nu_{\rm H}$ from Eq.~(\ref{2}), the Hall-viscosity induced correction to $R_{\rm H}$ near the ``universal" value of  $R_{\rm H}=-1/en_0c$ is written as
\be
R_{\rm H}\big\vert_{B=0}\simeq -\frac{1}{en_0c}\left(1+\frac{4\pi}{{\cal L}^2}\,n_dl_{\rm ee}^2\right)~.
\label{51}
\ee
That is, in the hydrodynamic regime, this is a small correction. It is, however, conceptually significant that Hall viscosity modifies, for the specular boundary condition, not only the dissipative resistivity, as discussed above, but also the Hall coefficient. Since the Hall-viscosity induced term in the Hall coefficient is a correction to the universal value, it is, arguably, more easily amenable to experimental investigation.

Note that the modification of $R_{\rm H}\big\vert_{B=0}$ by the {\it kinetic} coefficient $\nu_{\rm H}$, in particular, the presence of
the single relaxation rate $1/\tau_{\rm ee}$ (not a ratio of some relaxation rates, which may reduce to a quantity that by itself does not describe any kinetic process) in Eq.~(\ref{51}) contradicts the general statement from Ref.~\cite{auerbach18,*auerbach19} that $R_{\rm H}\big\vert_{B=0}$ is always expressible as a certain {\it thermodynamic} susceptibility, irrespective of the presence of disorder. While $R_{\rm H}$ in our framework explicitly depends on viscosity, with a clear origin of this dependence, the discrepancy is likely traceable to the decoupling procedure of disorder averaging (the handling of irreducible averages) at zero external momentum in Ref.~\cite{auerbach18,*auerbach19}.

\subsection{Dissipative resistivity}
\label{s9f}

Let us now discuss the dissipative resistivity in more detail. To order ${\cal O}(1/{\cal L})$, the momentum relaxation rate $1/\tau=\rho_{xx}\times(e^2n_0/m)$ that can be inferred from Eq.~(\ref{67a}) or, equivalently, from Eqs.~(\ref{64a}) and (\ref{62}), is given by
\be
\frac{1}{\tau}\simeq \frac{8\pi n_d}{\cal L}\,\nu~.
\label{78}
\ee
One remarkable point to note is that the electric field that defines $1/\tau$ in Eqs.~(\ref{64})-(\ref{65}) (which is in line with the above definition of $1/\tau$ in terms of $\rho_{xx}$) is the total electric field $\langle{\bf E}\rangle$, which includes the electric field inside obstacles, while the work performed by the electric field on the liquid per unit time and unit area is given by $-en_0\langle{\bf E}{\bf v}\rangle$, which, by definition, is entirely determined by the electric field in the liquid. This implies a subtle relation between the average product $\langle{\bf E}{\bf v}\rangle$ and the product of averages $\langle{\bf E}\rangle_{\rm liq}\langle{\bf v}\rangle$. Specifically, as follows from Eq.~(\ref{61a}),
\be
\langle{\bf E}{\bf v}\rangle\simeq 2\langle{\bf E}\rangle_{\rm liq}\langle{\bf v}\rangle~.
\label{79}
\ee
This relation can also be inferred from the correlation between the spatial behavior of $\nabla^2{\bf v}$ and $\bf v$ in the single-obstacle problem (Sec.~\ref{s7}).

Another point worth discussing in more detail is the viscosity-modulated magnetoresistance, especially in comparison with the magnetoresistance of noninteracting electrons. The momentum relaxation rate (\ref{78}) should be contrasted with $1/\tau=(8/3)n_dv_FR$ for impenetrable disks (in the quasiclassical limit of $mv_FR/\hbar\gg 1$) at zero $B$ in the absence of electron-electron interactions. Substituting Eq.~(\ref{1}) for $\nu$ in Eq.~(\ref{78}), the momentum relaxation is seen to become much weaker in the hydrodynamic regime of $\nu/{\cal L}\ll v_FR$, i.e., $v_F\tau_{\rm ee}/{\cal L}\ll R$, depending on the magnetic field and vanishing in the large-$B$ limit. The hydrodynamic lubrication (the decrease of $1/\tau$ as $\nu$ decreases) is thus enhanced by the magnetic field. This leads to a strong negative magnetoresistance $\Delta\rho_{xx}=\rho_{xx}(B)-\rho_{xx}(0)$, which for $\nu$ from Eq.~(\ref{1}) is representable to order ${\cal O}(1/{\cal L})$ as
\be
\frac{\Delta\rho_{xx}}{\rho_{xx}(0)}\simeq -\frac{(2\omega_c\tau_{\rm ee})^2}{1+(2\omega_c\tau_{\rm ee})^2}~.
\label{52}
\ee
The Lorentzian shape of $\rho_{xx}(B)$ in Eq.~(\ref{52}) implies that the hydrodynamic description of transport is valid for arbitrary $B$, including $B=0$ \cite{more_gen}.

Recall that $\Delta\rho_{xx}=0$ in the most conventional formulation of the Drude approach to transport of noninteracting electrons. In this formulation, among other conditions \cite{drude}, the magnetic field is assumed to not modify the scattering cross-section for momentum relaxation. Phenomenologically, the nonzero $\Delta\rho_{xx}$ in Eq.~(\ref{52}) could thus be simply the statement that the total-momentum relaxation rate varies with changing $B$. We emphasize, however, that the nonzero right-hand side of Eq.~(\ref{67c}) clearly signifies a departure from the Drude-like (characterizable solely and completely by the momentum relaxation rate) picture of magnetotransport \cite{drude}.

It is also worth noting that it is entirely because of electron-electron interactions that $\rho_{xx}$ does not exactly vanish above a critical magnetic field in the model of impenetrable obstacles (with no additional source of total-momentum relaxation). In this model, the Drude approach to magnetotransport of noninteracting electrons is totally inadequate \cite{baskin78,*baskin98}. The noninteracting kinetic problem for this model is exactly solvable \cite{bobylev95,*bobylev97} for classical electrons in the limit of $n_d\to\infty$ with $n_dR^2$ held fixed (``Boltzmann-Grad limit"), showing $\rho_{xx}\propto 1/B$ for large $B$. The metal-insulator transition mentioned above occurs beyond the Boltzmann-Grad limit at a critical value of the Larmor radius $R_c$ of the order of $n_d^{-1/2}$ \cite{baskin78,*baskin98,bobylev95,*bobylev97}. Therefore, one of the important messages of the hydrodynamic approach we developed here is that viscosity, induced by electron-electron interactions, restores
ergodicity of quasiclassical magnetotransport in the model of impenetrable obstacles.

\subsection{Drag and lift forces}
\label{s9e}

It is instructive to look at the emergence of $\Delta\omega_c$ [Eq.~(\ref{54})] from a different perspective. While the constant $C_1$, when averaged over different obstacles, determines the vector $\bf p$ [Eq.~(\ref{57})], and thus the resistivity tensor $\hat\rho$ [Eqs.~(\ref{21}) and (\ref{22})], it also defines the total force $\bf F$ exerted by the flowing liquid on a given obstacle. Specifically, the stress tensor $\sigma_{ij}$, which describes the distribution of forces in the liquid, is written in our problem as
\be
\sigma_{ij}=mn_0\nu\left[-\widetilde{\Omega}\delta_{ij}+(\partial_iv_j+\partial_jv_i)\right]~.
\label{68}
\ee
Here the pressure term, proportional to $\delta_{ij}$, includes, apart from the effect of $\nabla\phi$ on $\partial_t{\bf v}$ in Eq.~(\ref{3}), also the effect of both the Lorentz force and the Hall viscosity force [Eq.~(\ref{40a})], and the term in the round brackets gives the conventional contribution to $\sigma_{ij}$ of dissipative viscosity.

Imagine a closed contour in the liquid. The stress tensor defines the local force $\bf f$ per unit length of the contour exerted by the surrounding liquid on it as
\be
f_i=\sigma_{ij}N_j~,
\label{68a}
\ee
where $\bf N$ is the unit vector normal to the contour at the given point of it and oriented in the outward direction. The total force $\bf F$ acting on the obstacle is then given by the contour integral along its boundary
\be
{\bf F}=\oint_R\!dl\,\bf f~,
\label{69}
\ee
which, by using Eqs.~(\ref{68}) and (\ref{68a}), yields
\be
{\bf F}=mn_0\nu\left(\oint_R\!d{\bf l}\,\Omega+{\bf e}_z\times\oint_R\!d{\bf l}\,\widetilde\Omega\right)~^,
\label{72}
\ee
or, in terms of $C_1$ [Eq.~(\ref{71})],
\be
{\bf F}=-4\pi mn_0\nu\!\left(
\begin{array}{c}
{\rm Im}\\
{\rm Re}\\
\end{array}
\right)\!
C_1~.
\label{73}
\ee

Averaged over obstacles, $\langle{\bf F}\rangle$ is related to $\bf p$ [Eq.~(\ref{57})] as
\be
\langle{\bf F}\rangle=2\pi en_0{\bf p}
\label{74}
\ee
and can be split into two components,
\be
\langle{\bf F}\rangle=\langle{\bf F}\rangle_{\rm drag}+\langle{\bf F}\rangle_{\rm lift}~,
\label{75}
\ee
where the drag force $\langle{\bf F}\rangle_{\rm drag}$ is parallel to $\langle{\bf v}\rangle$ and the lift force $\langle{\bf F}\rangle_{\rm lift}$ is perpendicular to it. From Eqs.~(\ref{60a})-(\ref{61}), we have
\be
\langle{\bf F}\rangle_{\rm drag}\simeq \frac{8\pi mn_0\nu}{\cal L}\,\langle{\bf v}\rangle
\label{76}
\ee
and, by using Eq.~(\ref{54}) for $\Delta\omega_c$,
\be
\langle{\bf F}\rangle_{\rm lift}\simeq mn_0\,\frac{\Delta\omega_c}{n_d}\left(\langle{\bf v}\rangle\times{\bf n}\right)~.
\label{77}
\ee
Note that $-n_d\langle{\bf F}\rangle$ is the disorder-averaged force density exerted on the liquid by obstacles. Parenthetically, in the conventional hydrodynamic context, Eqs.~(\ref{72}) and (\ref{73}) describe also the force per unit length exerted on a cylinder immersed in a flow homogeneous along the axis of the cylinder (with $n_0$ understood then as the 3D density of the liquid).

As seen from Eq.~(\ref{77}), the emergence of $\Delta\omega_c$ is directly connected with the existence of the lift force exerted by the liquid on the obstacle (or rather the oppositely directed force exerted by the obstacle on the liquid). Note that, according to Eqs.~(\ref{25}), (\ref{41m}), and (\ref{71}) on the one hand and Eq.~(\ref{68}) on the other hand, half of the contribution to $\langle{\bf F}\rangle_{\rm lift}$ comes from the pressure and the other half from the viscous stress. The same relation holds for $\langle{\bf F}\rangle_{\rm drag}$ \cite{tomotika50}.

It is worth emphasizing that the lift force $\langle{\bf F}\rangle_{\rm lift}$ in Eq.~(\ref{77})---induced by Hall viscosity---is only due to the presence of $\nu_{\rm H}$ in the specular boundary condition but not the action of the Hall viscosity force in the bulk. This is in line with the proof given in Ref.~\cite{ganeshan17} that hydrodynamic lift is absent for the diffusive boundary condition, irrespective of the presence of Hall viscosity. Indeed, $\Delta\omega_c$ vanishes in the limit of $\nu_{\rm H}/\nu\to\infty$ [Eq.~(\ref{25a})] and thus in the case of the diffusive boundary condition.

It is perhaps also worthwhile to stress that the lift force in Eq.~(\ref{77}) is obtained in the linear-response limit, and so it is fundamentally different from inertial lift forces (both those of the ``aerodynamic" Kutta-Zhukovsky, or Joukowski, type \cite{lanlif6}, which emerge in inviscid fluids in the presence of velocity circulation around the obstacle, and those of the Saffman type \cite{saffman65,*saffman65corr}, which emerge in viscous fluids in the presence of simple shear). Viewed in a broader context, the lift force (\ref{77}) bears resemblance to the lift force characteristic of (incompressible) 2D active chiral liquids \cite{lou22}. It also shows similarity to the lift force exerted on a liquid domain by surrounding liquid when the two are characterized by different odd viscosity coefficients \cite{hosaka21a}.

\section{Flow-induced charge distribution}
\label{s8}

In Sec.~\ref{s7} and Appendix~\ref{a}, we discussed the behavior of the stream function $\psi({\bf r})$, the velocity ${\bf v}({\bf r})$, the vorticity $\Omega({\bf r})$, and the viscous force $m\nu\nabla^2{\bf v}({\bf r})$ for the flow past an obstacle. We now turn to the distribution of the charge density $-e[n({\bf r})-n_0]$ around the obstacle. As already mentioned below Eq.~(\ref{3}), we neglect the (small in the parameter $a_B/R$ \cite{compress} and not interesting here) term in the balance of forces that comes from a spatial variation of the degeneracy pressure $(\pi\hbar^2/m)n^2$ (per spin), i.e., the contribution to $\nabla\phi$ of the chemical potential. The electric potential $V$ is then given by $(m/e)\phi$ [Eq.~(\ref{3h})], and the charge density $-e(n-n_0)$ that creates the electric potential obeys the integral equation
\be
\phi({\bf r})=-\frac{e^2}{m}\int_{r'>R}\!d^2{\bf r}'\,\frac{n({\bf r}')-n_0}{|{\bf r}-{\bf r}'|}~.
\label{42}
\ee
We emphasize that $V$ and $n$ are related here through the bare (unscreened) Coulomb kernel. This is because $V$ and $n$ are the {\it actual} (screened) potential and density.

From Eqs.~(\ref{40a}) and (\ref{42}), the flow-induced density is seen to have two essentially different ingredients (this is, in effect, a continuation of the logic of Sec.~\ref{s3}),
\be
n-n_0=n_v+n_c~,
\label{42a}
\ee
where $n_v$, associated with the terms proportional to $\widetilde\Omega$ and $\Omega$ in $\phi$, arises because of viscosity and $n_c$, associated with the $\psi$ term in $\phi$, is the density induced directly by the cyclotron force. It is worth noting that the electric potential $(m/e)\phi$ is given by Eq.~(\ref{40a}) only inside the liquid but obeys Eq.~(\ref{42}) inside both the liquid and impenetrable obstacles.

For the stream function $\psi({\bf r})$ in the form of only the first angular harmonic, as is the case in the single-obstacle problem in Sec.~\ref{s7} or in the mean-field problem in Appendix~\ref{a}, the potential $\phi({\bf r})$ is of the same rotational symmetry. By expanding the Coulomb kernel in the 2D plane in cylindrical harmonics,
\be
\frac{1}{|{\bf r}-{\bf r}'|}=\sum_{m=-\infty}^\infty\!\!e^{im(\varphi-\varphi')}\int_0^\infty\!\!dk\,J_m(kr)J_m(kr')~,
\label{85b}
\ee
where $J_m(x)$ is the Bessel function of the first kind, doing the Hankel transform, and using the identity
\be
\int_0^\infty\!\!dk\,kJ_m(kr)J_m(kr')=\frac{1}{r}\delta(r-r')~,
\label{86b}
\ee
the solution to Eq.~(\ref{42}) for $n({\bf r})$ can then be written as
\be
{\rm N}(r)=-\frac{m}{e^2}\int_0^\infty\!\!dk\,k^2J_1(kr)\int_0^\infty\!\!dr'r'{\rm F}(r')J_1(kr')~,
\label{87b}
\ee
with ${\rm N}(r)$ and ${\rm F}(r)$ being the amplitudes of the angular harmonics of $n({\bf r})-n_0={\rm Re}\left\{{\rm N}(r)e^{i\varphi}\right\}$
and $\phi({\bf r})={\rm Re}\left\{{\rm F}(r)e^{i\varphi}\right\}$, respectively.

The solution (\ref{87b}), while being exact, implies the knowledge of $\phi({\bf r})$ in the whole 2D plane. In particular, the problem of finding the charge distribution induced by a flow inside the circle with $r<L$ in the model consideration of Sec.~\ref{s7} is well posed (has a unique solution) only if one specifies whether there are ``external" (for $r>L$) charges and, if yes, what are then the boundary conditions for the electrostatic problem. However, for $L/R\gg 1$, any physically reasonable boundary condition imposed (at the model level) on $n({\bf r})$ and/or $\phi({\bf r})$ in the area with $r>L$ only slightly affects the dipole part of the charge distribution around the obstacle for $r\ll L$ [as can also be inferred from Eq.~(\ref{87b})]. Alternatively, one can rely on the disorder-averaged solution for the flow around a given obstacle, the position of which is fixed, in a random obstacle array (Appendix~\ref{a}). Below, to describe the density distribution around the obstacle, we primarily make use of the solution of the single-obstacle problem from Sec.~\ref{s7}. We focus on the viscosity-induced charge density $n_v(\bf{r})$ and its connection with the notion of the Landauer dipole in Secs.~\ref{s8a}--\ref{s8c}, while relegating the discussion of $n_c({\bf r})$ to Appendix \ref{b}.

\subsection{Viscosity-induced dipole}
\label{s8a}

Solving Eq.~(\ref{42}) for $n_v$ with $\widetilde\Omega$ and $\Omega$ from Eqs.~(\ref{24a}) and (\ref{70}) asymptotically for $R\ll r\ll L$, we have
\be
n_v\simeq \frac{m}{\pi e^2r^2}\,{\rm Im}\left\{C_1(\nu+is\nu_{\rm H})e^{i\varphi}\right\}~.
\label{43}
\ee
Equation (\ref{43}) is obtained by noting that
\be
\int\!d^2{\bf r}'\,\frac{{\bf r}{\bf r}'}{(r')^3|{\bf r}-{\bf r}'|}=2\pi~,
\label{44}
\ee
where the integration is performed over the whole space, with $n_v({\bf r})$ inversely proportional to the integral. As already mentioned in Sec.~\ref{s3} [Eq.~(\ref{3d2})] and now demonstrated in Eq.~(\ref{43}), the viscosity-related charge density in the 2D electron liquid falls off away from the obstacle in the limit of ideal screening as $1/r^2$.

It is worth stressing that Eq.~(\ref{43}) for $n_v$ is purely classical, with the prefactor $1/\pi e^2$  which is expressible as $2a_B\partial n/\partial\mu$ in terms of two quantum quantities, the Bohr radius $a_B$ (screening radius in a degenerate electron gas) and the inverse compressibility $(\partial n/\partial\mu)^{-1}$. The limit of ideal screening, in which we neglected the variation of the chemical potential in $\nabla\phi$, means the limit $a_B/R\to 0$ with the product $a_B\partial n/\partial\mu$ held fixed \cite{compress}.

For $L/R\gg 1$, the dependence of $n_v({\bf r})$ on the length scales $R$ and $L$ in Eq.~(\ref{43}) is only under the sign of the logarithm in $C_1$ [Eq.~(\ref{35a})]. In a random array of obstacles, for the viscosity-induced density $\overline{n}_v({\bf r})$ around a given obstacle averaged over positions of other obstacles (Sec.~\ref{s9a} and Appendix~\ref{a}), the logarithmic dependence is specified as
\be
\overline{n}_v\simeq -\frac{2m}{\pi e^2{\cal L}\,r^2}\,|\langle{\bf v}\rangle|\,{\rm Re}\left\{(\nu+is\nu_{\rm H})e^{i\left(\varphi-\varphi_{\langle{\bf v}\rangle}\right)}\right\}
\label{49}
\ee
for $R\ll r\ll L_v$. Since the density $n_v({\bf r})$ in Eq.~(\ref{43}) is sharply peaked at the obstacle, the compact dipoles induced by the flow in typical realizations of the ensemble of rare obstacles can be viewed as separate entities, each associated with its own obstacle.

It is notable that the dipole axis of the distribution of $\overline{n}_v({\bf r})$ in Eq.~(\ref{49}) is rotated by Hall viscosity: the axis is parallel to $\langle{\bf v}\rangle$ for $\nu_{\rm H}=0$ and perpendicular to it in the limit of large $\nu_{\rm H}/\nu$, irrespective of the degree of specularity in the boundary condition. The angle $\theta$ by which the direction of the electric dipole in Eq.~(\ref{49}) is rotated from the direction of $\langle{\bf v}\rangle$ is given, for ${\cal L}\gg 1$, by
\be
\theta\simeq -\arctan\frac{s\nu_H}{\nu}
\label{44a}
\ee
(the sign $\simeq$ only refers to the shape of the dependence of $\theta$ on $B$, with $\theta$ varying between the exact values of $\theta=0$ at $B=0$ and $\theta=\pm\pi/2$ in the large-$B$ limit).

Equation (\ref{49}) describes the leading contribution to $\overline{n}_v({\bf r})$, which does not depend on the boundary condition. An additional Hall-viscosity induced rotation of the dipole axis emerges in the expansion of $\overline{n}_v({\bf r})$ in powers of $1/{\cal L}$ at order ${\cal O}(1/{\cal L}^2)$ for the case of the specular boundary condition. Importantly, the density distribution is affected by Hall viscosity even if the velocity distribution is not, which is the case for the diffusive boundary condition (Secs.~\ref{s4} and \ref{s5}), and indeed it is the redistribution of $n({\bf r})$ with varying $\nu_{\rm H}$ that is necessary to maintain the lack of dependence of ${\bf v}({\bf r})$ on $\nu_H$ in that case. Measuring the charge distribution induced by the flow around an obstacle provides thus a direct way to probe Hall viscosity.

Note that both $\overline{n}_v({\bf r})$ in Eq.~(\ref{49}) and $\overline{n}_c({\bf r})$ in Eq.~(\ref{47}) have a dipole structure. However, generically, the dipoles in $\overline{n}_c({\bf r})$ and $\overline{n}_v({\bf r})$ are oriented differently: the dipole orientation of $\overline{n}_v({\bf r})$ is highly sensitive to the presence of Hall viscosity [Eq.~(\ref{44a})], whereas that of $\overline{n}_c({\bf r})$ does not depend on $\nu_{\rm H}/\nu$ in the leading approximation for ${\cal L}\gg 1$. The dipoles in $\overline{n}_v({\bf r})$ and $\overline{n}_c({\bf r})$ are oriented along the same axis (in opposite directions), perpendicular to $\langle{\bf v}\rangle$, only in the limit of a strong magnetic field \cite{reuss96remark,reuss96}. The amplitude of the charge modulation around the obstacle also depends strongly on the magnetic field. For $B=0$, there is only $\overline{n}_v$ present. It is only in the large-$B$ limit, namely for the Larmor radius $R_c\ll R$, that the amplitude of the variation of $\overline{n}_c({\bf r})$ with changing $\varphi$ becomes [$(R/R_c)^2$ times] larger than that of $\overline{n}_v({\bf r})$. Note also that, irrespective of the strength of the magnetic field, the density $\overline{n}_v({\bf r})$ is sharply peaked at the obstacle, whereas the density $\overline{n}_c({\bf r})$ is spread on the scale of $L_v\gg R$.

\subsection{Electric polarization and Hall viscosity}
\label{s8b}

Notice that the angle $\theta$ in Eq.~(\ref{44a}) is the same angle by which the electric field {\it inside} obstacles $\langle{\bf E}\rangle_{\rm obs}$ is rotated by Hall viscosity in Eq.~(\ref{n4}). Recall that the angle by which (the bulk contribution to) the electric field $\langle{\bf E}\rangle_{\rm liq}$ inside the liquid---outside the obstacles---is rotated by Hall viscosity in Eq.~(\ref{41d}) comes with the opposite sign.
As already mentioned below Eq.~(\ref{41h}), the two Hall-viscosity induced contributions to the total field $\langle{\bf E}\rangle$ cancel each other exactly. In the presence of Hall viscosity, we thus encounter a nontrivial distribution of the electric field. As a matter of fact, as we will see below, the problem is not completely trivial even for $\nu_{\rm H}=0$.

In the limit of rare obstacles, the viscosity-induced electric field inside an obstacle can be thought of as being produced only by charges forming the dipole (\ref{43}) around the same obstacle. Note that the charge distribution described by Eq.~(\ref{43}) [or Eq.~(\ref{49}) for that matter] is not characterized by a finite dipole moment: since the dipolar charge density falls off as $1/r^2$, the dipole moment diverges (linearly with $r$). With this in mind, it is convenient to use the exact mapping expressed by Eq.~(\ref{3d1}).

As discussed in Sec.~\ref{s3} [see also the comment below Eq.~(\ref{77})], the distribution of forces exerted by obstacles on the liquid in the $(x,y)$ plane is exactly the same for the 2D flow through an array of impenetrable disks and the 3D flow through an array of impenetrable parallel cylinders oriented along the $z$ axis. In particular, the distribution of the flow-induced electric field within the plane is identical in the two systems, only the flow-induced charge density differs. Recall that, in the limit of ideal screening, charges induced by an arbitrary 3D flow for $B=0$ are only present on the surface of obstacles [Eq.~(\ref{3a})]. Because of the homogeneity of the system of parallel cylinders in the $z$ direction, this is the case with regard to viscosity-induced charges also for $B\neq 0$ if the magnetic field is oriented along the cylinders.
The averaged electric field inside obstacles for the 2D flow [Eq.~(\ref{n4})] can thus by represented as a polarization field:
\be
\langle{\bf E}\rangle_{\rm obs}=-2\pi n_d{\bf q}_{\rm cyl}~,
\label{80}
\ee
where
\be
{\bf q}_{\rm cyl}=\frac{m}{2\pi e}\,{\bf e}_z\times\left\langle\oint_{ob}\!d{\bf l}\,\left(\nu\widetilde\Omega-s\nu_{\rm H}\Omega\right)\right\rangle
\label{81}
\ee
is the averaged dipole moment per unit length of a cylinder. This dipole moment is strongly affected by Hall viscosity, being rotated by the angle $\theta$ [Eq.~(\ref{44a})] from the direction of $\langle{\bf v}\rangle$.

Since
\be
\left\langle\oint_{ob}\!d{\bf l}\,\Omega\right\rangle={\bf e}_z\times\left\langle\oint_{ob}\!d{\bf l}\,\widetilde\Omega\right\rangle
\ee
in the limit of rare obstacles [recall the comment below Eq.~(\ref{71})], the vector $\bf p$ [Eq.~(\ref{22a})], which determines the total electric field $\langle{\bf E}\rangle$, can thus be expanded in this limit as
a sum of four terms:
\be
{\bf p}=({\bf q}_1+{\bf q}_2)+({\bf q}_1-{\bf q}_2)~,
\label{82}
\ee
where ${\bf q}_1$ and ${\bf q}_2$ are given by the first and second terms in Eq.~(\ref{81}), respectively. The Hall-viscosity induced contribution ${\bf q}_2$ cancels out from $\bf p$, with
\be
{\bf p}=2{\bf q}_1
\label{83}
\ee
being twice as large as the term ${\bf q}_1$ in the dipole ${\bf q}_{\rm cyl}$. The factor of 2 in Eq.~(\ref{83}) is essentially the manifestation of the ``fifty-fifty split" in Eq.~(\ref{61a}). To dispel a possible confusion: despite the cancellation of ${\bf q}_2\propto\nu_{\rm H}$ from $\bf p$, the remaining part of ${\bf p}$ in Eq.~(\ref{82}) depends on $\nu_{\rm H}$ for the specular boundary condition, and this is how Hall viscosity affects then the resistivity tensor $\hat\rho$ (Sec.~\ref{s9}).

\subsection{Landauer dipole in 2D hydrodynamics}
\label{s8c}

The notion of the average force acting on electrons $-e\langle{\bf E}\rangle$ being determined in viscous plasma hydrodynamics by electric polarization of obstacles, as formalized by Eqs.~(\ref{22x})-(\ref{22}) and (\ref{83}), parallels that of a ``Landauer dipole" \cite{landauer57,*landauer75,sorbello81,*sorbello98,sorbello88,zwerger91,kunze95}. This charge density perturbation is formed around a scatterer in a conducting electron system by elastically scattered electron waves weighted with the nonequilibrium, describing a nonzero current, distribution function. The Coulomb interaction only manifests itself in the Landauer-dipole theory through static screening of the current-induced electric potential. In the hydrodynamic limit, considered here, fast electron-electron interactions modify the picture of a Landauer dipole in an essential way; however, the dipolar structure of the nonequilibrium electron density around the obstacle remains intact (Sec.~\ref{s8a}), together with Landauer's idea of the current-induced electric field in a conductor with rare scatterers being extremely inhomogeneous.

It is worth emphasizing that the charge density perturbation conventionally designated ``the Landauer dipole" is {\it not} the actual charge
density around a scatterer. Rather, it is the bare (unscreened) density obtainable within the noninteracting picture of scattered waves. In the 2D case, the bare and actual densities fall off with increasing distance $r$ to the scatterer as $1/r$ \cite{sorbello88,zwerger91} and $1/r^2$, respectively. In the formulation of our hydrodynamic problem, the density $n({\bf r})$ is, throughout the paper, the actual density.

Let us now delve deeper into the question of how Eq.~(\ref{22x}) for $\langle{\bf E}\rangle$ and Eq.~(\ref{80}) for $\langle{\bf E}\rangle_{\rm obs}$, with $\bf p$ and ${\bf q}_{\rm cyl}$ related by Eq.~(\ref{83}), are connected with the original idea \cite{landauer57,*landauer75} that the average electric field $\langle{\bf E}\rangle$ in a 3D conductor for $B=0$ is expressible in terms of the average current-induced dipole moment at an impurity ${\bf p}_{\rm 3D}$ as $\langle{\bf E}\rangle=-4\pi n_d{\bf p}_{\rm 3D}$, with $n_d$ understood as the 3D impurity density. This equation reproduces the Drude formula for noninteracting electrons. A direct extension of the reasoning \cite{landauer57,*landauer75} to the case of noninteracting 2D electrons scattered at $B=0$ off impurities the 2D density of which is $n_d$ yields \cite{sorbello88}
\be
\langle{\bf E}\rangle=-4\pi n_d{\bf q}~,
\label{84}
\ee
where the Landauer dipole moment $\bf q$ is defined in terms of the electric potential
\be
V({\bf r})=2\frac{{\bf q}{\bf r}}{r^2}
\label{85}
\ee
created by scattered electrons at nonequilibrium around an impurity \cite{factoroftwo}. The electric field for $V({\bf r})$ from Eq.~(\ref{85}) is given by
\be
{\bf E}({\bf r})=\frac{2}{r^2}\left[-{\bf q}+2\frac{({\bf q}{\bf r}){\bf r}}{r^2}\right]_{r>0+}\!\!\!\!-2\pi{\bf q}\,\delta({\bf r})~,
\label{85a}
\ee
where the last term describes the polarization field in the ``core" of the dipole.

The relation (\ref{84}) corresponds to the resistance measurement between the source and drain terminals in the form of two infinitely long parallel lines in the 2D plane \cite{sorbello88}. The condition on the geometry of the area over which the electric field is averaged is crucially important here, with the right-hand side of Eq.~(\ref{84}) being a sum of two equal contributions. Specifically, one of them comes from the polarization field of the dipole and the other from the averaging of the ``stray" field [the term in the square brackets in Eq.~(\ref{85a})] produced by the dipole on the spatial scale given by the distance between the source and drain lines.

In the absence of Hall viscosity, there is thus a one-to-one correspondence in terms of phenomenology between the expressions for $\langle{\bf E}\rangle$ in Eqs.~(\ref{22x}) (hydrodynamics) and (\ref{84}) (Drude theory). Namely, ${\bf p}$ in Eq.~(\ref{22x}) is given by $2{\bf q}_{\rm cyl}$, with ${\bf q}_{\rm cyl}$ being equivalent to ${\bf q}$ in Eq.~(\ref{84}). However, as discussed in Sec.~\ref{s8b}, for $\nu_{\rm H}\neq 0$, the direct generalization of the Landauer dipole to hydrodynamics ${\bf q}_{\rm cyl}={\bf q}_1+{\bf q}_2$ contains the Hall-viscosity induced component ${\bf q}_2$, while the vector $\bf p$, determining $\langle{\bf E}\rangle$, is given by $2{\bf q}_1$ [Eq.~(\ref{83})] irrespective of the presence or absence of Hall viscosity.

At this point, it is useful to look at a simple---but highly instructive in the context of the hydrodynamic Landauer dipole---electrostatic problem of finding the average electric field within a finite 2D system of identical pointlike ``dipoles" each creating the potential (\ref{85}) (we use here the quotation marks since this potential is created not by a 2D dipole as such but by a cylinder with the dipole density $\bf q$ perpendicular to the plane \cite{factoroftwo}). The average electric field $\langle{\bf E}\rangle$ within the system is a sum of the polarization fields inside the dipole cores and the average field $\langle{\bf E}\rangle_{\rm out}$ outside them. For illustrative purpose, assume that the dipoles form a regular (square) lattice of a rectangular shape, altogether $L\times M$ dipoles sitting at $x=(1,2,\ldots,L)a$ and $y=(1,2,\ldots,M)a$, where $a$ is the lattice constant. Let  the area of averaging be a rectangle with opposite vertices at the points $(x,y)$ fixed as $(0,0)$ and $[(L+1)a,(M+1)a]$. The contribution to $\langle{\bf E}\rangle a^2$ of the polarization fields is given by $-2\pi {\bf q}$ and the contribution of the field in the area outside the pointlike dipoles by
\begin{align}
&\langle{\bf E}\rangle_{\rm out} a^2=2\pi{\bf q}-\frac{8}{LM}\nonumber\\
&\times\left({\bf e}_xq_x\sum_{n=1}^L\sum_{m=1}^M+\,{\bf e}_yq_y\sum_{n=1}^M\sum_{m=1}^L\right)\arctan \frac{m}{n}~,
\label{86}
\end{align}
where ${\bf e}_x$ and ${\bf e}_y$ are the unit vectors in the $x$ and $y$ directions.
For $\bf q$ oriented along the $x$ axis, we have
\be
\langle{\bf E}\rangle a^2\simeq -4{\bf q}\,\frac{M}{L}\ln\frac{L}{M}
\label{87}
\ee
for $L\gg M\gg 1$ and
\be
\langle{\bf E}\rangle a^2\simeq -4\pi {\bf q}
\label{88}
\ee
for $M\gg L\gg 1$. By symmetry, $\langle{\bf E}\rangle_{\rm out}=0$ for $L=M$.

Equation (\ref{88}) is the most relevant to the notion of the Landauer dipole, as it shows explicitly the origin of the doubling of $\langle{\bf E}\rangle$ in Eq.~(\ref{84}) compared to the average polarization field. For $M\gg L$, the current flows across a narrow stripe, so that the polarization and stray fields produced inside the stripe by the Landauer dipoles contribute, as already mentioned below Eq.~(\ref{85a}), equally to $\langle{\bf E}\rangle$. By contrast, if the dipoles are oriented along the narrow stripe, $\langle{\bf E}\rangle$ in Eq.~(\ref{87}) is seen to vanish in the limit of a large aspect ratio $L/M$. This is not what happens if the current flows along the narrow stripe (Hall-bar geometry) with impurities because the electric field in this geometry is mainly produced by the current-induced charges that are {\it additional} to those forming the Landauer dipoles (see the discussion at the very end of Sec.~\ref{s6aB}). The geometry with $M\gg L$ is peculiar precisely with regard to the fact that only in this limit the charges produced by the boundaries of the sample do not play a role in the distribution of the electric field.

One of the most remarkable features of the hydrodynamic expression for the average electric field $\langle{\bf E}\rangle$ obtained in the thermodynamic limit [Eq.~(\ref{41k})] is that it is valid---as already emphasized at the very end of Sec.~\ref{s6aA}---for an arbitrary shape of the sample. In particular, Eq.~(\ref{22x}) holds irrespective of the aspect ratio of the sample containing a macroscopically large number of obstacles. In view of the discussion below Eq.~(\ref{88}), this means that the hydrodynamic transport theory developed in Sec.~\ref{s6a} ``automatically" incorporates the production of charges whose  density varies on the scale of the system size. The subtle point here is that the average electric field $\langle{\bf E}\rangle$ in the thermodynamic limit is ``universally" determined---regardless of the presence or absence of these additional charges---by the average contribution of an individual obstacle, as formalized by Eq.~(\ref{22x}) in terms of the vector $\bf p$.

Within the concept of the Landauer dipole, the fact that $\langle{\bf E}\rangle$ is determined in Eqs.~(\ref{22x}), (\ref{82}), and (\ref{83}) by only the component ${\bf q}_1$ of the flow-induced dipole (\ref{81}), whereas the Hall-viscosity induced component ${\bf q}_2$ cancels out from $\langle{\bf E}\rangle$, is rationalized by means of the auxiliary electrostatic model presented above. This model demonstrates that the cancellation implies a very specific choice of the sample geometry when using Landauer's line of approach. Specifically, the sample should be chosen in the form of a stripe with $M/L\to\infty$ oriented normally to the vector $\bf p$. If ${\bf q}_2\neq 0$, the Landauer dipole ${\bf q}_{\rm cyl}$ is then not perpendicular to the axis of the stripe, and its component ${\bf q}_2$ is ``filtered out" by the averaging over the long stripe. Indeed, as Eq.~(\ref{86}) shows, $\langle{\bf E}\rangle a^2\to -4\pi q_x{\bf e}_x$ in the limit of $M\gg L\gg 1$. Crucially, in the presence of Hall viscosity and for the case of the specular boundary condition, the vector ${\bf p}$ is oriented at an angle with respect to the average velocity $\langle{\bf v}\rangle$. This fundamentally limits the universality of Landauer's formulation as far as the conceptual link between the (dissipative) resistivity and the Landauer dipole in hydrodynamics is concerned.

\section{Experiment}
\label{s10}

Before concluding, we briefly comment on the possible relevance of the hydrodynamic description of magnetotransport in a random array of impenetrable obstacles to high-mobility GaAs heterostructures. We are primarily concerned with these structures because there have been, for quite some time, experimental indications pointing towards the importance of rare strong scatterers in them. More specifically, these indications point towards the interplay of strong scatterers and smooth disorder---both at zero and, especially, at a classically strong magnetic field, including linear-response transport and nonequilibrium phenomena, for an early review see Ref.~\cite{dmitriev12}.

One particularly significant set of experiments, which might be thought of as potentially relevant to our discussion above, are the measurements of $\rho_{xx}$ as a function of $B$ (in sufficiently wide samples to exclude the effect of their boundaries on $\rho_{xx}$). A very strong negative magnetoresistance, with $\rho_{xx}$ dropping by a factor of up to several tens \cite{longresasym}, was reported for $B$ of a fraction of kG and $T$ down to a few hundred mK \cite{dai10,bockhorn11,hatke12,mani13,bockhorn14,shi14a,nmr,hatke11,dai11} (see also Ref.~\cite{wang22} for the wider samples among those studied there). The magnetoresistance is weakly $T$ dependent in the low-$T$ limit but strongly suppressed for $T$ above 1-1.5\,K in ultrahigh-mobility samples \cite{bockhorn11,dai11,hatke12,mani13,bockhorn14,wang22} or persisting to $T$ an order of magnitude higher in the case of moderate mobility (with $\rho_{xx}$ dropping by a factor of about ten) \cite{shi14a}. Changes in the negative magnetoresistance in ultrahigh-mobility samples brought about by addition of an artificially created random array of impenetrable obstacles were studied---for various densities of the array---in Ref.~\cite{horn-cosfeld21}. The phenomenon is likely of a classical origin (the ultrahigh mobility of the order of $10^7\,{\rm cm}^2/{\rm V\,s}$ is too high to otherwise explain the amplitude of $\Delta\rho_{xx}$).

The type of strong scatterers that is commonly thought to limit electron mobility in ultrahigh-mobility GaAs heterostructures is ``background impurities" (as opposed to ``remote donors") present in a small concentration in close vicinity of the 2D electron system, see Ref.~\cite{umansky09} and Sec.~II.A in Ref.~\cite{dmitriev12}. It is important to note that the model of Refs.~\cite{baskin78,*baskin98,bobylev95,*bobylev97}, mentioned at the end of Sec.~\ref{s9f}, is entirely inadequate to describe magnetotransport controlled by background impurities. This is because of the presence of smooth disorder produced by remote donors, the result of which is that the typical shift of the guiding center of a cyclotron orbit after one revolution is, for the magnetic fields at which the experiments \cite{dai10,bockhorn11,hatke12,mani13,bockhorn14,shi14a,nmr,hatke11,dai11} were performed, by far larger than the characteristic size of background impurities (which is the Bohr radius) \cite{dmitriev12}.

Thinking more generally, however, reveals that correlations in the multiple scattering process at the points of self-intersection of diffusing quasiclassical paths produce a strong negative magnetoresistance \cite{mirlin01} for the case of ``mixed disorder" (strong scatterers plus smooth disorder), with the magnetic field effectively ``switching off" scattering off strong scatterers. These correlations are key to the concept of quasiclassical ``memory effects" and display very rich physics, especially in magnetotransport \cite{dmitriev12}. Adding electron-electron scattering, which by itself conserves total momentum, to the picture of quasiclassical correlations will provide a $T$ dependence of the magnetoresistance induced by the memory effects \cite{memory+ee}. With this in mind, it remains to be seen if this theoretical framework is capable of shedding more light on the experimental observations \cite{dai10,bockhorn11,hatke12,mani13,bockhorn14,shi14a,nmr,hatke11,dai11,horn-cosfeld21}, in particular, with regard to the discussions of the possible connection between the two in Refs.~\cite{dai10,hatke12,shi14a,bockhorn14,horn-cosfeld21}.

In the last decade or so, it has become clear \cite{bockhorn14} that a different type of strong scatterers may be relevant to transport in high-mobility GaAs heterostructures, namely Ga droplets (``oval defects"). These typically emerge in the technological process in very small numbers (with the characteristic density $10^4\,{\rm cm}^{-2}$) but can be very large in size (with the characteristic radius up to about $20\,\mu{\rm m}$ inside the 2D electron system, as reported in Ref.~\cite{bockhorn14}). Although these droplets were commonly deemed irrelevant to the transport measurements, impenetrable obstacles of that size and density put a bound on the electron mean free path $l_{\rm P}$ at a scale comparable to $l_{\rm P}\sim 10^2\,\mu{\rm m}$ typically deducible from the mobility in ultrahigh-mobility samples. Moreover, they seem to have ``come to the fore" in the literature in what concerns the memory effects of the type \cite{mirlin01,bockhorn14} and the measured magnetoresistance.

Note that memory effects of a similar nature may be present simultaneously for both types of strong scatterers (background impurities and Ga droplets), thus giving rise to two peaks in the shape of the dependence of $\rho_{xx}$ on $B$, both centered at $B=0$, superimposed on each other. The characteristic amplitude and width of the two peaks may substantially differ in view of the differences in the characteristic concentration and radius for the two types of strong scatterers.

There is difficulty, however, in reliably analyzing the results of Refs.~\cite{dai10,bockhorn11,hatke12,mani13,bockhorn14,shi14a,nmr,hatke11,dai11}
in these terms: (at least) the size of Ga droplets is apparently very sensitive to the technological process. Specifically, the Ga droplets in the ``best wafers" grown in the Weizmann Institute are (at most) of the same density $10^4\,{\rm cm}^{-2}$ as in the ultrahigh-mobility samples on which the measurements were performed in Ref.~\cite{bockhorn14}---but their characteristic radius is an order of magnitude smaller, about $1\,\mu{\rm m}$ \cite{umansky_private}. The density of Ga droplets can also vary noticeably \cite{bockhorn16} depending on the growth conditions. From this perspective, it might be interesting to look into how the results of the magnetotransport measurements correlate with the amount and size of Ga droplets.

Besides intentionally created large obstacles (e.g., the obstacle radius in Ref.~\cite{horn-cosfeld21} was fixed at $0.5\,\mu{\rm m}$), Ga droplets may thus be one of the experimentally realizable examples of impenetrable obstacles of the type discussed in the present paper. It should be noted, however, that although it might be tempting to view the viscous magnetoresistance described in the present paper as directly related to the observations in Refs.~\cite{dai10,bockhorn11,hatke12,mani13,bockhorn14,shi14a,nmr,hatke11,dai11,horn-cosfeld21}, the hydrodynamic picture for transport in the bulk should be taken with caution as far as the electron liquid in high-mobility GaAs samples for $T\sim 1\,$K or less is concerned.

Indeed, the characteristic $l_{\rm ee}$ for these samples at $T=1$\,K is likely about 100\,$\mu$m for $B=0$ \cite{l_ee}, which gives $\tau_{\rm ee}=l_{\rm ee}/v_F$ of the same order of magnitude as the total-momentum relaxation time $\tau$ for ultrahigh-mobility samples, where $\tau$ is extracted from the measured mobility $\mu$ according to $\tau=(m/e)\mu$. In the moderate mobility sample from Ref.~\cite{shi14a}, in which a very strong negative magnetoresistance was observed, the thus obtained $\tau$ is, correspondingly, an order of magnitude smaller than $\tau_{\rm ee}$.
The condition $\tau\ll\tau_{\rm ee}$ also implies that the total-momentum relaxation length is much smaller than $l_{\rm ee}$ \cite{size_eff}.
The conventional hydrodynamic approach, assuming that viscosity is induced by electron-electron interactions, is thus unlikely to be justified at zero $B$ for transport in the bulk of moderate- to ultrahigh-mobility GaAs structures---all the more so because $l_{\rm ee}$ exceeds the size of strong scatterers, even if these are associated with Ga droplets. In view of the latter condition, transport at $B=0$ is likely describable in the spirit of Refs.~\cite{guo16,shytov18}, with the addition of the effect of smooth disorder.

Turning to the observed dependence of $\rho_{xx}$ on $B$ \cite{dai10,bockhorn11,hatke12,mani13,bockhorn14,shi14a,nmr,hatke11,dai11,horn-cosfeld21}, which is often referred to in the literature, following Ref.~\cite{alekseev16}, as a manifestation of viscous hydrodynamics in macroscopic 2D samples, an essential inconsistency between the experimental data and its description in terms of viscous transport looms. Within the hydrodynamic framework, the negative magnetoresistance results from the suppression of dissipative viscosity according to Eq.~(\ref{1}), as discussed in Sec.~\ref{s9f}. To articulate the issue in general terms, let us change over to phenomenology by introducing the relaxation time of the second angular harmonic of the distribution function $\tau_2$ (which reduces to $\tau_{\rm ee}$ in the case of viscosity induced by electron-electron scattering). Extracting $\tau_2$ from the measured width of the peak of $\rho_{xx}(B)$, which is invariably a fraction of kG in high-mobility GaAs heterostructures, the thus obtained $\tau_2$ at $T$ about 1\,K is conspicuously by far too small, giving the length $l_2=v_F\tau_2$ of the order of a few microns. In particular, if one assumes that $\tau_2$ is given by $\tau_{\rm ee}$, this length scale is at least an order of magnitude less than expected.

In fact, the observed $T$ dependence of the width of the peak in $\rho_{xx}(B)$ saturates with decreasing $T$ in this range of $T$, which indicates---within the linear-response hydrodynamic picture---that $l_2$ in the low-$T$ limit is not given by $l_{\rm ee}$ but, rather, is connected with electron scattering off disorder. The condition of applicability of the hydrodynamic framework to describe viscous flow around impenetrable obstacles of radius $R$ would then be $l_2\ll R$, which requires the presence of an additional source of disorder, on top of impenetrable obstacles. In turn, the purported predominately hydrodynamic character of the flow would require that the viscosity-related total-momentum relaxation
time $\tau$, given by Eq.~(\ref{64a}), be much smaller than the total-momentum relaxation time $l_{\rm P}/v_F$ induced by the additional source of disorder. These conditions are hardly compatible with each other in the discussed framework, so that viscosity induced by impurity scattering is unlikely to play any significant role in a good conductor with rare obstacles.

In terms of $l_2$ and $l_{\rm P}$ both induced by disorder, the essential problem with the description of the experimental results for the magnetoresistance within the picture of viscous hydrodynamics thus manifests itself in the fact that fitting to the experimental data in the low-$T$ limit within this picture would require that $l_2$ be much smaller, by a large margin (at least an order of magnitude), than $l_{\rm P}$. Note also that, although a strong magnetic field makes the hydrodynamic regime more easily achievable \cite{more_gen}, this line of approach implies the existence of an intermediate, as $B$ increases, regime in which a strong magnetoresistance emerges not describable in terms of conventional hydrodynamics, but electron-electron interactions in the presence of disorder play a crucial role.

The above means that there is much difficulty in associating the magnetoresistance in Refs.~\cite{dai10,bockhorn11,hatke12,mani13,bockhorn14,shi14a,nmr,hatke11,dai11,horn-cosfeld21,wang22} directly with the purely hydrodynamic description of the electron liquid flowing past strong scatterers.
To summarize, (i) the characteristic $\tau_{\rm ee}$ is too large to account for a hydrodynamic regime for $B=0$, (ii) the characteristic $\tau_2$ extracted from the measured width of the peak of $\rho_{xx}$ as a function of $B$ under the assumption of viscous transport is by far too small, (iii) the observed saturation of the width of the peak of $\rho_{xx}$ with decreasing $T$ cannot possibly be associated with disorder-induced viscosity. It is worth mentioning that this cautionary remark is at odds with the forthright conclusion made in Ref.~\cite{alekseev16}. As a matter of fact, the fitting in Ref.~\cite{alekseev16} to the experimental data \cite{shi14a} (for the sample parameters cited in Ref.~\cite{shi14a} with the mobility and electron density at $T=1\,$K about $1.0\times 10^6\,{\rm cm}^2/{\rm V\,s}$ and $2.8\times 10^{11}\,{\rm cm}^{-2}$, respectively) yields the fitting parameters that are rather dubious precisely with regard to the inconsistency outlined above. In particular, the lengths $l_2$ and $l_{\rm P}$ used in the fitting (Fig.~1 in Ref.~\cite{alekseev16}) at $T=1\,$K are $3\,\mu{\rm m}$ and $100\,\mu{\rm m}$, respectively, differing by more than an order of magnitude, with the length scale of $3\,\mu{\rm m}$ being thus much smaller than both $l_{\rm ee}$ for $T=1\,$K and $l_{\rm P}$. Moreover, the ``effective sample width" $10\,\mu{\rm m}$, used in the fitting and associated with the average distance between Ga droplets with reference to the experimental data in
Ref.~\cite{bockhorn14}, is an order of magnitude smaller than about $90\,\mu{\rm m}$ corresponding to the droplet density
$1.3\times 10^4\,{\rm cm}^{-2}$ cited in Ref.~\cite{bockhorn14}.

Altogether, despite the above note of caution, viscous hydrodynamics is an important direction in attempting to understand the magnetoresistance observed in Refs.~\cite{dai10,bockhorn11,hatke12,mani13,bockhorn14,shi14a,nmr,hatke11,dai11,horn-cosfeld21}, with the present paper formulating the theoretical framework for electron hydrodynamics in a random array of impenetrable obstacles in the presence of a magnetic field and solving the basic problem in this direction.

\section{Summary}
\label{s11}

We have presented a detailed analysis of the flow of the 2D electron liquid through a random ensemble of rare impenetrable obstacles in the presence of a magnetic field. The theoretical framework we have formulated to calculate the linear-response resistivity tensor $\hat\rho$ relates $\hat\rho$ to the vorticity and its harmonic conjugate, both averaged along the boundaries of the obstacles. This basic relation shows that $\hat\rho$, which is defined by the averaged electric field induced by the electron flow, has two distinctly different contributions. One is related to the electric field induced in the liquid, the other to the electric field induced inside obstacles. Remarkably, the electric fields outside and inside obstacles give equal contributions to the dissipative resistivity $\rho_{xx}$ in the limit of rare obstacles. This, in particular, highlights an inherent link between hydrodynamics and electrostatics in the charged liquid. Specifically, the contribution to $\rho_{xx}$ of the electric field in the liquid is brought about by viscous stress, whereas that of the electric field inside obstacles comes from pressure exerted by them on the liquid.

Throughout the paper, we have maintained the theme of elucidating the role played by Hall viscosity in transport of the electron liquid past obstacles. We have shown that the averaged electric fields outside and inside obstacles are rotated by Hall viscosity from the direction of the averaged velocity. For the diffusive boundary condition on the obstacles, this effect exactly cancels in $\hat\rho$, as a result of which $\hat\rho$ is not affected by Hall viscosity. By contrast, the total electric field is modified by Hall viscosity for the specular boundary condition. One conceptually interesting piece of physics here is that the resulting dependence of $\hat\rho$ on Hall viscosity implies the emergence of an effective---proportional to the obstacle density---magnetic field produced by Hall viscosity in addition to the external one. The dependence of the Hall resistivity $\rho_{xy}$ on Hall viscosity leads to a deviation of the Hall constant from its universal value.

Within the analytically controllable approach, we have also described the vanishing of $\rho_{xx}$ in the limit of a strong magnetic field, the essential physics of which is the known modification of the dissipative viscosity coefficient by the magnetic field. A corollary of the magnetic-field enhanced hydrodynamic lubrication is that hydrodynamics can, in principle, be probed in magnetotransport through a random array of rare obstacles in the bulk of a sample.

We have further explored the interplay of hydrodynamics and electrostatics by calculating the distribution of charges that create the flow-induced electric field around obstacles. We have related the resistivity in the hydrodynamic regime with the disorder-averaged electric dipole induced by viscosity at the obstacle. This conceptual link is much in the spirit of the Landauer dipole for noninteracting particles, with the dipole precisely defined in our problem in terms of hydrodynamic variables. We have also shown that the viscosity-induced dipole is rotated from the flow direction by Hall viscosity.

Although it might be tantalizing to speculate that the enhancement of the hydrodynamic lubrication by the magnetic field is behind the numerous fairly puzzling observations of the strong negative magnetoresistance in high-mobility GaAs heterostructures, the experimental picture is likely far from being described in purely hydrodynamic terms, as we have also discussed in the paper.

{\it Note added.}---Very recently, when the writing of this manuscript was nearing completion, the preprint \cite{alekseev23} has appeared dealing with some of the problems studied in our work.

\acknowledgments

We thank P.~S. Alekseev and A.~P. Dmitriev for useful discussions. We are also grateful to L.~Bockhorn, R.~J. Haug, J.~H. Smet, and M.~A. Zudov for valuable discussions concerning the details of the experiments. We are especially thankful to V. Umansky for an in-depth discussion regarding certain aspects of the sample characterization. This work was supported by the European Commission under the EU Horizon 2020 MSCA-RISE-2019 Program (Project No.\ 873028 HYDROTRONICS) and by the German-Israeli Foundation for Scientific Research and Development (GIF) Grant No.\ I-1505-303.10/201.

\appendix

\section{MEAN-FIELD FORMULATION}
\label{a}

As mentioned below Eqs.~(\ref{67a})-(\ref{67c}), a complementary way to organize the calculation of $\hat\rho$, which gives the same results and has the same level of accuracy with respect to the expansion of $\hat\rho$ in powers of $1/{\cal L}$, is the mean-field approximation (``effective medium approximation"). The advantage of this approach is that it gives an explicit form for the hydrodynamic variables around an obstacle the position of which is fixed, averaged over positions of other obstacles. Below, the quantities that are subjected to this type of disorder averaging are denoted by a bar over them.

The essence of the mean-field approximation is to ensure the balance of disorder-averaged forces in the linearized Navier-Stokes equation around a given obstacle by representing the effect of other obstacles as friction and lift forces, both local in space. Specifically, the velocity $\overline{\bf v}$ in the stationary case obeys [cf.\ Eq.~(\ref{3})]
\be
\nabla\overline\phi-\omega_c(\overline{\bf v}\times{\bf n})+\nu\nabla^2\overline{\bf v}-\nu_{\rm H}(\nabla^2\overline{\bf v}\times{\bf n})+I(\overline{\bf v})=0~,
\label{a1}
\ee
where the term $I(\overline{\bf v})$ is local and linear in $\overline{\bf v}$:
\be
I(\overline{\bf v})=-\overline{\bf v}/\tau-\Delta\omega_c(\overline{\bf v}\times{\bf n})~.
\label{a2}
\ee
In Eq.~(\ref{a2}), the total-momentum relaxation rate $1/\tau$ and the effective cyclotron frequency $\Delta\omega_c$, which describes hydrodynamic lift for the specular boundary condition, are given by Eqs.~(\ref{78}) [see also Eq.~(\ref{64a})] and (\ref{54}), respectively.
Equation (\ref{a1}) is supplemented with the condition
\be
\nabla\overline{\bf v}=0
\label{a3}
\ee
[cf.\ Eq.~(\ref{2})]. The boundary conditions to Eqs.~(\ref{a1}) and (\ref{a3}) are fixed on the boundary of the given obstacle at $r=R$ [Eqs.~(\ref{4})-(\ref{6})] and at infinity as
\be
\overline{\bf v}\vert_{r\to\infty}=\langle{\bf v}\rangle~,
\label{a4}
\ee
where $\langle{\bf v}\rangle$ is defined in Eq.~(\ref{12e2}).

It may be worth noting that the force density $mn_0I(\overline{\bf v})$---exerted by obstacles (except the one that does not participate in disorder averaging) on the liquid---counterbalances on average the force $\bf F$ [Eq.~(\ref{72})] for each of the obstacles and, as such, has contributions of the electric field both inside the liquid and inside the obstacles [cf.\ Eq.~(\ref{72}) and the averaging of the electric field in Sec.~\ref{s6a}]. Indeed, recall that, for any given obstacle, the part of $\bf F$ stemming from pressure [the second term in Eq.~(\ref{72})] is representable for the charged liquid ``rushing against" the rigid walls of the obstacle in terms of the electric field inside the obstacle. Accordingly, $(m/e)I(\overline{\bf v})$ for $r\to\infty$ is given by $\langle{\bf E}\rangle-\langle{\bf E}_{\rm H}\rangle$, where $\langle{\bf E}\rangle$ is the total field [Eq.~(\ref{n3})]. Note also that the balance of forces in Eq.~(\ref{a1}) is the stationary limit of the dynamic equation with the term $\partial_t\overline{\bf v}$ on the right-hand side of Eq.~(\ref{a1}), which describes dynamics of the momentum density carried by the liquid. The contribution to $\partial_t\overline{\bf v}$ from the friction force is given by $-\overline{\bf v}/\tau$, where $1/\tau$ is the momentum relaxation rate that determines the resistivity, i.e., once more, the total---including the contributions from the area outside and inside obstacles---electric field $\langle{\bf E}\rangle$ induced by the current.

The mean-field approach to Stokes flow in a random 3D array of obstacles was conceptualized in the late 1940s \cite{brinkman49}, with the mean-field equation similar to Eq.~(\ref{a1}) derived for the 3D case for $\omega_c=0$ in the absence of Hall viscosity and for the diffusive boundary condition (hence with $\Delta\omega_c=0$) \cite{tam69,childress72,lundgren72,howells74,rubinstein86}. It was explicitly demonstrated that the mean-field approximation in the 3D problem is exact in the limit of a dilute array to linear order in the density of obstacles.

An important feature of the 2D model is that the 2D resistivity at linear order in $n_d$ is expandable in a series in powers of $1/{\cal L}$ (Sec.~\ref{s9}), as was also noted \cite{howells74} within the hierarchical scheme of decoupling multiple-obstacle ``interactions" \cite{childress72} when it is applied to the 2D case. In the mean-field framework, the parameter that controls this expansion is represented as $1/n_dL_v^2\ll 1$, where the relaxation length $L_v$ (for the decay of a viscous effect produced on the flow by a given obstacle) is given by Eq.~(\ref{62}). That is, firstly, it is the large number of obstacles within the ``relaxation area" $L_v^2$ that justifies the mean-field approximation to leading order in $1/{\cal L}$. Recall that the leading order is ${\cal O}(1/{\cal L})$ for $\rho_{xx,0}$ and ${\cal O}(1/{\cal L}^2)$ for both $\rho_{xx,{\rm H}}$ and $s\rho_{xy}-m\omega_c/e^2n_0$ [cf.\ Eqs.~(\ref{67a})-(\ref{67c})]. Secondly, in contrast to the 3D case, spatial fluctuations of the number and the positions of obstacles within the area $L_v^2$, which amounts to spatial fluctuations of the relaxation length, break the mean-field approximation in the form of Eqs.~(\ref{a1}) and (\ref{a2}) for the calculation of $\hat\rho$ in the 2D case already at linear order in $n_d$. This occurs at the first subleading order in $1/{\cal L}$ beyond the leading one \cite{remark_mf}.

The mean-field equation (\ref{a1}) formalizes essentially a single-obstacle problem, similar to that solved in Sec.~\ref{s7}, with the outer boundary moved to infinity [cf.\ Eqs.~(\ref{9}) and (\ref{a4})] and the starting equation for $\psi$ modified [cf.\ Eq.~(\ref{11})] as
\be
\nabla^4\overline\psi-\frac{1}{L_v^2}\nabla^2\overline\psi=0~.
\label{a5}
\ee
The solution to Eq.~(\ref{a5}) for the boundary conditions specified above is the first angular harmonic $\overline\psi({\bf r})={\rm Re}\left\{\overline\chi(r)e^{i\varphi}\right\}$ [cf.\ Eq.~(\ref{12})] with the function $\overline\chi$ obeying Eqs.~(\ref{16}) or (\ref{17a}) on the boundary of the obstacle and
\be
\left(\overline\chi/r\right)\vert_{r\to\infty}= -i|\langle{\bf v}\rangle|e^{-i\varphi_{\langle{\bf v}\rangle}}
\label{a6}
\ee
at infinity. The solution reads
\be
\overline\chi=D_1\frac{r}{R}+D_2\frac{R}{r}+D_3K_1\!\left(\frac{r}{L_v}\right)~,
\label{a7}
\ee
where $K_m(x)$ is the modified Bessel function of the second kind, the coefficients $D_1$ and $D_2$ are exactly given by $D_1=-iR|\langle{\bf v}\rangle|e^{-i\varphi_{\langle{\bf v}\rangle}}$ and $D_2= -D_1-D_3K_1(R/L_v)$, and the coefficient $D_3$ to leading order in $L_v/R$ [with relative corrections to $D_3/L_v$ of the order of ${\cal O}(R^2/L_v^2)$] is given, for the specular boundary condition, by
\be
D_3\simeq -\frac{2iL_v}{\ln(L_v/R)-h/2 +d}\,|\langle{\bf v}\rangle|e^{-i\varphi_{\langle{\bf v}\rangle}}~.
\label{a8}
\ee
Here $d=\ln 2+1/2-\gamma$, with $\gamma$ being Euler's constant, and $h$ is defined by Eq.~(\ref{34a}). As in Secs.~\ref{s7} and \ref{s9}, the result for the diffusive boundary condition is obtainable by sending $\nu_{\rm H}/\nu\to\infty$ ($h\to 1$) [Eq.~(\ref{25a})]. Note that the first two terms in $\overline\chi$ in Eq.~(\ref{a7}) drop out from $\overline\Omega$ and $\nabla^2\overline{\bf v}$ (cf.\ the corresponding terms in $\chi$ in Sec.~\ref{s7}).

As discussed above, similarly to the calculation of $\hat\rho$ in Sec.~\ref{s9}, retaining the $\nu_{\rm H}$ independent term $d$ in the denominator of Eq.~(\ref{a8}) is beyond the accuracy of the mean-field approximation. This is in contrast to the term $-h/2$ that accurately describes the leading---of the order of ${\cal O}(1/{\cal L}^2)$---contributions to the Hall-viscosity induced terms in $\hat\rho$ [cf.\ the denominator of $C_1$ in Eq.~(\ref{35a})]. By representing, similarly to Eqs.~ (\ref{60b}) and (\ref{60a}), $\overline\Omega$ and $\overline{\bf v}$ as
$\overline{\Omega}=\overline{\Omega}_0+\overline{\Omega}_{\rm H}$ and $\overline{\bf v}=\overline{\bf v}_0+\overline{\bf v}_{\rm H}$, where the first terms in the sums do not depend on $\nu_{\rm H}$ and the second terms are induced by Hall viscosity, we have from Eqs.~(\ref{a7}) and (\ref{a8}):
\begin{align}
&\overline{\Omega}_0\simeq -\frac{4\,|\langle{\bf v}\rangle|}{L_v{\cal L}}\,
K_1\!\left(\frac{r}{L_v}\right)\sin\left(\varphi-\varphi_{\langle{\bf v}\rangle}\right)~,
\label{a9}\\
&\overline{\Omega}_{\rm H}\simeq -\frac{4\,|\langle{\bf v}\rangle|}{L_v{\cal L}^2}\,
K_1\!\left(\frac{r}{L_v}\right){\rm Im}\left\{he^{i\left(\varphi-\varphi_{\langle{\bf v}\rangle}\right)}\right\}
\label{a10}
\end{align}
for the leading terms in $\overline\Omega_{0,{\rm H}}$, and
\begin{align}
\nabla^2\overline{\bf v}_0\simeq \frac{2|\langle{\bf v}\rangle|}{L_v^2{\cal L}}
\!\left(
\begin{array}{c}
{\rm Re}\\
{\rm Im}\\
\end{array}
\right)&\!\left\{\left[\,K_2\!\left(\frac{r}{L_v}\right)e^{2i\left(\varphi-i\varphi_{\langle{\bf v}\rangle}\right)}
\right.\right.\nonumber\\
&\left.\left.\,\,-K_0\!\left(\frac{r}{L_v}\right)\,\right]e^{i\varphi_{\langle{\bf v}\rangle}}\right\}~,
\label{a11}\\
\nabla^2\overline{\bf v}_{\rm H}\simeq \frac{2|\langle{\bf v}\rangle|}{L_v^2{\cal L}^2}
\!\left(
\begin{array}{c}
{\rm Re}\\
{\rm Im}\\
\end{array}
\right)&\!\left\{\left[\,hK_2\!\left(\frac{r}{L_v}\right)e^{2i\left(\varphi-i\varphi_{\langle{\bf v}\rangle}\right)}
\right.\right.\nonumber\\
&\left.\left.\,\,-h^*K_0\!\left(\frac{r}{L_v}\right)\,\right]e^{i\varphi_{\langle{\bf v}\rangle}}\right\}
\label{a12}
\end{align}
for the leading terms in $\nabla^2\overline{\bf v}_{0,{\rm H}}$.

The amplitudes of the angular harmonics of $\overline{\bf v}$, defined in terms of $\overline\chi$ similarly to Eqs.~(\ref{26}) and (\ref{27a}), are given by
\begin{align}
\overline{g}_+&=  \left[\,1-\frac{1}{W}K_0\!\left(\frac{r}{L_v}\right)\,\right]|\langle{\bf v}\rangle|e^{-i\varphi_{\langle{\bf v}\rangle}}~,
\label{a13}\\
\overline{g}_-&= -\left\{\frac{R^2}{r^2}+\frac{1}{W}\left[\,\frac{2RL_v}{r^2}K_1\!\left(\frac{R}{L_v}\right)-K_2\!\left(\frac{r}{L_v}\right)\,\right]\right\}\nonumber\\
&\times|\langle{\bf v}\rangle|e^{-i\varphi_{\langle{\bf v}\rangle}}~,
\label{a14}
\end{align}
where
\be
W=K_0\!\left(\frac{R}{L_v}\right)+(1-h)\frac{R}{2L_v}K_1\!\left(\frac{R}{L_v}\right)~.
\label{a15}
\ee
This is the exact solution to Eq.~(\ref{a1}). We keep the term proportional to $K_1(R/L_v)$ in the square brackets in Eq.~(\ref{a14}) not expanded in the small parameter $R/L_v$ so that $\overline{\bf v}$ in Eqs.~(\ref{a13}) and (\ref{a14}) satisfies the boundary condition of impenetrability, $v_r=0$ at $r=R$, exactly. This condition is then satisfied for arbitrary $W$. Similarly, the exact form of $W$ guarantees that the other boundary condition at $r=R$ [Eq.~(\ref{6})] is also satisfied exactly (note that the difference between the specular and diffusive boundary conditions is fully encoded in $W$, with the second term in $W$ vanishing for the diffusive boundary condition). Otherwise, $\overline{g}_\pm$ from Eqs.~(\ref{a13})-(\ref{a15}) can be written to linear order in $n_dR^2$, expanded in powers of $1/{\cal L}$, and split into two parts, one describing $\overline{\bf v}_0$ and the other $\overline{\bf v}_{\rm H}$, each of which is within the limits of applicability of the mean-field approximation similarly to Eqs.~(\ref{a9})-(\ref{a12}).

Importantly, the terms in Eqs.~(\ref{a13}) and (\ref{a14}) at $r\sim L_v$ reproduce the ``power counting" presented between Eqs.~(\ref{57}) and (\ref{60b}) with regard to the expansion of the amplitude and the phase (for complex $h$) of $\overline{g}_\pm(L_v)$ in powers of $1/{\cal L}$. In particular, the mean-field approach produces, as follows from Eqs.~(\ref{a9}) and (\ref{a10}) at $r=R$, exactly the same leading terms in ${\bf p}_0$ [of the order of ${\cal O}(1/{\cal L})$] and ${\bf p}_{\rm H}$ [of the order of ${\cal O}(1/{\cal L}^2)$] as in Eqs.~(\ref{60}) and (\ref{61}) [with the subleading terms in the expansion of $\bf p$ in powers of $1/{\cal L}$ being, as already mentioned below Eq.~(\ref{a4}), beyond the accuracy of the mean-field approximation itself]. To order ${\cal O}(1/{\cal L})$, the drag force in Eq.~(\ref{76}) and the one obtained \cite{howells74} for the 2D case within the hierarchical scheme \cite{childress72} (for the diffusive boundary condition) coincide with each other.

It is worth noting the difference in the effect of disorder on the vorticity $\overline\Omega$ and the viscous force, proportional to $\nabla^2\overline{\bf v}$, on the one hand, and on the velocity $\overline{\bf v}$ itself on the other hand. Both $\overline\Omega$ and $\nabla^2\overline{\bf v}$ decay exponentially, as the distance from the given obstacle $r$ increases, on the scale of the relaxation length $L_v$, whereas the perturbation $\overline{\bf v}-\langle{\bf v}\rangle$ induced by the obstacle decays only as a power law. Specifically, the amplitude of the second angular harmonic of $\overline{\bf v}$ behaves, as follows from Eq.~(\ref{a14}) upon substitution of Eq.~(\ref{62}) for $L_v$, in the limit of $r\gg L_v$ as
\be
\overline{g}_-\simeq -\frac{1}{2\pi n_dr^2}|\langle{\bf v}\rangle|e^{-i\varphi_{\langle{\bf v}\rangle}}~,\qquad r\gg L_v
\label{a16}
\ee
[the contribution to $\overline{g}_-$ of Hall viscosity for $r\gg L_v$ is given by the expression in Eq.~(\ref{a16}) multiplied by the small factor $h/{\cal L}$]. Note that this is in contrast to the amplitude of the zeroth angular harmonic of $\overline{\bf v}-\langle{\bf v}\rangle$ which falls off with increasing $r$ proportionally to $\exp(-r/L_v)$ [Eq.~(\ref{a13})].

Another point to notice is that the equation obeyed by $\overline{g}_+$ from Eq.~(\ref{a13}),
\be
-\nabla^2\overline{g}_++\frac{1}{L_v^2}\left(\overline{g}_+-\overline{g}_+\vert_{r\to\infty}\right)=0~,
\label{a17}
\ee
incorporates the balance between the zeroth angular harmonics of the viscous force $m\nu\nabla^2\overline{\bf v}$ and the friction force $-m(\overline{\bf v}-\langle{\bf v}\rangle)/\tau$ that makes $\overline{\bf v}$ relax to $\langle{\bf v}\rangle$:
\be
\int_0^{2\pi}\!
\!d\varphi\left(\nabla^2\overline{\bf v}-\frac{\overline{\bf v}-\langle{\bf v}\rangle}{L_v^2}\right)=0~.
\label{a18}
\ee
It follows that the sum of the zeroth harmonics of $\nabla^2\overline{\bf v}$ and $-\overline{\bf v}/L_v^2$, in which each of the two terms is rotated with respect to $\langle{\bf v}\rangle$ because of Hall viscosity, is oriented strictly along $\langle{\bf v}\rangle$. In turn, the component of the zeroth harmonic of the disorder-averaged pressure term $\nabla\overline{\phi}$ in Eq.~(\ref{a1}) that counterbalances $\nu\nabla^2\overline{\bf v}-\overline{\bf v}/\tau$ is constant and given by $\langle{\bf v}\rangle/\tau$ for arbitrary $r$. This is in contrast to the component of the zeroth harmonic of $\nabla\overline\phi$ that is due to Hall viscosity, which varies in space as a linear function of $\overline{\bf v}$.

The balance of forces in Eq.~(\ref{a18}) demonstrates also a subtle effect produced by the environment of a given obstacle on the behavior of the viscous force $m\nu\nabla^2\overline{\bf v}$ around it on spatial scales below the relaxation length $L_v$. Specifically, as $r$ increases, the amplitude of the zeroth harmonic of $\nabla^2\overline{\bf v}$ decreases logarithmically for $R\leq r\ll L_v$ as $\ln(L_v/r)$ [Eqs.~(\ref{a11}) and (\ref{a12})]. This counterbalances the corresponding change of the force $-m(\overline{\bf v}-\langle{\bf v}\rangle)/\tau$. The logarithmic relaxation of the zeroth harmonic of $\nabla^2\overline{\bf v}$ should be contrasted with Eq.~(\ref{24b}) for the single-obstacle problem, where the zeroth harmonic of $\nabla^2{\bf v}$ does not depend on $r$. Note that the second harmonic of $\nabla^2\overline{\bf v}$ in Eqs.~(\ref{a11}) and (\ref{a12}) varies in the limit of $r\ll L_v$ as $1/r^2$, which is, in contrast to the zeroth harmonic, in exact correspondence with the behavior of the second harmonic of $\nabla^2{\bf v}$ in Eq.~(\ref{24b}). Correspondingly, the vorticity $\overline\Omega$ in Eqs.~(\ref{a9}) and (\ref{a10}) consists for $r\ll L_v$ of two leading terms, one of which falls off with increasing $r$ as $1/r$, exactly as in Eq.~(\ref{24a}) for the single-obstacle problem, whereas the other grows as $r\ln(L_v/r)$, with the additional logarithmically varying factor compared to Eq.~(\ref{24a}). As already mentioned above, the result for $\overline\Omega$ at $r=R$ reproduces exactly the leading terms in both ${\bf p}_0$ and ${\bf p}_{\rm H}$ in Eqs.~(\ref{60}) and (\ref{61}).

Finally, we should note that equations similar to Eqs.~(\ref{a1}) and (\ref{a2}) can be written outside the context of the mean-field approximation for a random array of obstacles. Namely, as mentioned at the beginning of Sec.~\ref{s7}, this is the case in a single-obstacle problem if the regularizator of Stokes' singularity is a total-momentum relaxation [finite $1/\tau$ in Eq.~(\ref{a2})] produced by weak ``environmental" disorder, which can be static or dynamic, around the obstacle. From this perspective, the key distinguishing property of the mean-field approximation in a random obstacle array is the ``self-consistency" condition on the total-momentum relaxation rate in Eq.~(\ref{a2}), which in our case takes the form of Eqs.~(\ref{64}) and (\ref{64a}) in the limit of ${\cal L}\gg 1$. Note that equations of the Navier-Stokes type for a single obstacle with environment-induced friction in the presence of odd viscosity arise \cite{banerjee17,lou22} in the context of active chiral liquids, where they become similar to Eqs.~(\ref{a1}) and (\ref{a2}) if one neglects rotational viscosity. As already mentioned at the end of Sec.~\ref{s5}, the specular boundary condition is then the same in the active chiral liquid and in the magnetized electron liquid.

\section{LORENTZ-FORCE INDUCED DIPOLE}
\label{b}

In Secs.~\ref{s8a} and \ref{s8b}, we discussed the flow-induced perturbation of the electron density around an obstacle $n_v({\bf r})$ that is directly produced by viscosity forces (associated with both dissipative and Hall viscosities). In contrast to $n_v({\bf r})$, the Lorentz-force induced density $n_c({\bf r})$ is not peaked in the vicinity of the obstacle. Indeed, the variation of the Hall electric field $-(m\omega_c/e)({\bf v}\times{\bf n})$ around the obstacle follows the slowing down of the flow around it, so that its amplitude behaves only logarithmically as a function of $r$. Specifically, according to Eqs.~(\ref{33a}) and (\ref{36})---apart from the intricate behavior, associated with Hall viscosity, in the immediate vicinity of the obstacle on the scale of $r\sim R$---the amplitude of the Hall field grows slowly as $\ln (r/R)$ with increasing $r$ for $R\ll r\ll L$. This requires that the charges producing the variation of the Hall field with $r$ be spread on the scale of $L$ [not $R$, as in the case of $n_v({\bf r})$].

Solving Eq.~(\ref{42}) for $n_c({\bf r})$ in the limit of $R\ll r\ll L$, we have
\begin{align}
n_c\simeq\frac{m\omega_c}{\pi e^2\ln(L/R)}\,\frac{{\bf r}}{r}\left({\bf v}_0\times{\bf n}\right)
\label{46}
\end{align}
for the charge density that produces the component of the Hall electric field that behaves as $\ln(L/r)$ on top of the homogeneous component $-(m\omega_c/e)({\bf v}_0\times{\bf n})$. The charge distribution (\ref{46}) forms a dipole around the obstacle and is $r$ independent. For the density $\overline{n}_c$ around a given obstacle, averaged over positions of other obstacles (Sec.~\ref{s9a} and Appendix \ref{a}), Eq.~(\ref{46}) is modified to
\be
\overline{n}_c\simeq\frac{2m\omega_c}{\pi e^2{\cal L}}\,\frac{{\bf r}}{r}\left(\langle{\bf v}\rangle\times{\bf n}\right)
\label{47}
\ee
for $R\ll r\ll L_v$. Similarly to Eq.~(\ref{49}) for $\overline{n}_v$, Eq.~(\ref{47}) gives the leading contribution to $\overline{n}_c$ in the expansion of $\overline{n}_c$ in powers of $1/{\cal L}$.

It is instructive to examine how $n_c({\bf r})$ behaves beyond the limit of $r\ll L$ when a boundary condition, mimicking the effect of other obstacles around a given one, is placed on $r>L$ within the model problem of Sec.~\ref{s7}. A simple and illustrative solution can be obtained for the condition that specifies the potential $\phi$, fixed as $\phi=-s\omega_c\psi$ with $\psi$ given by Eqs.~(\ref{12}) and (\ref{19b}) for $R<r<L$ and obeying $\nabla^2\phi=0$ for $r<R$ and $r>L$ with continuous $\nabla\phi$ at $r=R$ and $r=L$, which thus imposes no constraint on $n_c({\bf r})$ directly. The condition for $r\geq L$ means that $\Omega({\bf r})=0$ [Eq.~(\ref{11a})] and ${\bf v}({\bf r})={\bf v}_0$ for all $r\geq L$.

This problem is exactly solvable for $n_c({\bf r})$ in terms of $\Omega({\bf r})$:
\be
n_c({\bf r})=s\frac{m\omega_c}{4\pi^2e^2}\!\int_{R<r'<L}\!\!d^2{\bf r}'\,\frac{\Omega({\bf r}')}{|{\bf r}-{\bf r}'|}
\label{46a}
\ee
with $\Omega({\bf r})$ from Eq.~(\ref{24a}) and the same Coulomb kernel as in the ``direct" transformation in Eq.~(\ref{42}). In the limit of $L/R\gg 1$, Eq.~(\ref{46a}) yields $n_c({\bf r})$ for $r\gg R$ that differs from that in Eq.~(\ref{46}) by a factor $\kappa(r/L)$, where the dimensionless function $\kappa(x)$, expressible in terms of the Bessel functions as
\be
\kappa(x)=1-\int_0^\infty\!\frac{dk}{k}\,[J_0(k)+2J_2(k)]\,J_1(kx)~,
\label{46b}
\ee
describes the behavior of $n_c({\bf r})$ on the scale of $r\sim L$, with $\kappa(0)=1$ and $\kappa(\infty)=0$. This demonstrates in what way the vorticity in the presence of a magnetic field [Eqs.~(\ref{11a}) and (\ref{46a})] establishes a profile of $n_c({\bf r})$ that maintains the balance between the slowing down and acceleration of the flow around an obstacle in Eq.~(\ref{39}).

Extending the line of approach that led to Eqs.~(\ref{46a}) and (\ref{46b}) to the mean-field solution from Appendix~\ref{a}, $\overline{n}_c({\bf r})$ in Eq.~(\ref{47}) acquires the factor $\kappa_{\rm mf}(r/L_v)$, where
\be
\kappa_{\rm mf}(x)=\int_0^\infty\!dk\,\frac{k}{k^2+1}J_1(kx)~.
\label{46c}
\ee
The function $\kappa_{\rm mf}(r/L_v)$ specifies the crossover behavior of $\overline{n}_c({\bf r})$ for $r\sim L_v$ with the limiting values $\kappa_{\rm mf}(0+)=1$ and $\kappa_{\rm mf}(\infty)=0$.

\bibliography{viscmr}

\begin{thebibliography}{159}
\bibitem [{2vs()}]{2vs3}%
  \BibitemOpen
  \href@noop {} {}\bibinfo {note} {Strictly speaking, the dependence of the
  viscosity tensor (both the dissipative and Hall components) on $B$ was
  obtained in Refs.~\cite{chapman70,kaufman60,braginskii65,*braginskii58} for
  the case of three dimensions and in Ref.~\cite{alekseev16} for the case of
  two. However, the difference here only concerns the expression of $l_{\rm
  ee}$ in terms of the microscopic parameters of the system, while the
  dependence of the corresponding components of the viscosity tensor on $B$ for
  given $l_{\rm ee}$ is identical in both cases, as it should.}\BibitemShut
  {Stop}%
\bibitem [{sig()}]{sign_hallvisc}%
  \BibitemOpen
  \href@noop {} {}\bibinfo {note} {While the sign of $\nu$ must be positive for
  thermodynamic reasons, the sign of $\nu_{\rm H}$ is a matter of choice: we
  define it to be positive for any orientation of the magnetic
  field.}\BibitemShut {Stop}%
\bibitem [{thr()}]{three_d}%
  \BibitemOpen
  \href@noop {} {}\bibinfo {note} {In contrast to the simple form of
  Eq.~(\ref{3}), the flow in a 3D incompressible isotropic plasma placed in a
  magnetic field is characterized by five viscosity coefficients
  \cite{lanlif10,chapman70,kaufman60,braginskii65,*braginskii58}. At $B=0$,
  however, the Navier-Stokes equation is the same in the 2D and 3D cases, with
  $\nabla$ being of the corresponding dimensionality.}\BibitemShut {Stop}%
\bibitem [{edg()}]{edge}%
  \BibitemOpen
  \href@noop {} {}\bibinfo {note} {Pressure applied to a truly incompressible
  liquid with $\nu=0$ in the presence of $\nu_{\rm H}\neq 0$ (and/or
  $\omega_c\neq 0$) is countered by a dissipationless current along the edges
  (which may change the shape of the liquid, if the obstacle is not a hard
  wall) but makes no effect in the bulk.}\BibitemShut {Stop}%
\bibitem [{scr()}]{screened}%
  \BibitemOpen
  \href@noop {} {}\bibinfo {note} {The conductivity tensor $\hat\rho^{-1}$
  describes, by the conventional definition, the current as a response to the
  {\it screened} electric field (more precisely, to the gradient of the
  electrochemical potential; this difference is not important in the limit of
  zero screening radius \cite{compress}), which is the field that enters the
  Navier-Stokes equation.}\BibitemShut {Stop}%
\bibitem [{dru()}]{drude}%
  \BibitemOpen
  \href@noop {} {}\bibinfo {note} {In the Drude approach, based---in the first
  place---on the assumption that consecutive scattering events are a Markovian
  process, $s\rho_{xy}$ is exactly given by $m\omega_c/e^2n$ for a single type
  of particles with an isotropic dispersion relation if one can neglect the
  energy dependence of the momentum relaxation rate. The exact vanishing of
  $\Delta\rho_{xx}$ requires additionally that this rate be $B$
  independent.}\BibitemShut {Stop}%
\bibitem [{reu()}]{reuss96remark}%
  \BibitemOpen
  \href@noop {} {}\bibinfo {note} {A polarization bias by the Hall electric
  field of the charge density profile around an impurity added to the 2D
  electron gas occupying fully the lowest Landau level was discussed in terms
  of a ``Landauer resistivity dipole in a strong magnetic field" in
  Ref.~\cite{reuss96}.}\BibitemShut {Stop}%
\bibitem [{lon()}]{longresasym}%
  \BibitemOpen
  \href@noop {} {}\bibinfo {note} {Experimental data for the four-terminal
  longitudinal resistance show sometimes noticeable asymmetry for $B\to -B$,
  which urges caution when dealing with the cases of extremely strong negative
  magnetoresistance, such as the one for sample A in Fig. 4 in
  Ref.~\cite{dai10}, where the reported ratio $\rho_{xx}(B)/\rho_{xx}(0)$
  reaches the value of about 1/50. Specifically, if the longitudinal resistance
  were identified as the dissipative resistance and the ratio
  $\rho_{xx}(B)/\rho_{xx}(0)\ll 1$ were extracted from the data for only one
  sign of $B$, the result for given $|B|$ could differ substantially from that
  for the other sign of $B$ (which would mean, then, a clash with the Onsager
  relation). For example, strong asymmetry of this kind can be seen for $|B|$
  above about 0.4\,kG in Fig. 2 in Ref.~\cite{wang22}. The asymmetry for $B\to
  -B$ of the longitudinal resistance possibly means the admixture of
  $\rho_{xy}$ to $\rho_{xx}$ in the measured signal. One of the reasons for
  this may be the presence of an electron density gradient in the Hall bar
  \cite{ilan06,zhou15}.}\BibitemShut {Stop}%
\bibitem [{nmr()}]{nmr}%
  \BibitemOpen
  \href@noop {} {}\bibinfo {note} {Although the main focus of
  Refs.~\cite{hatke11,dai11} was on the photoresistance, they also showed data
  for a very strong negative magnetoresistance in the absence of
  illumination.}\BibitemShut {Stop}%
\bibitem [{l_e()}]{l_ee}%
  \BibitemOpen
  \href@noop {} {}\bibinfo {note} {This characteristic length scale for $l_{\rm
  ee}$ follows from a straightforward estimate within the random-phase
  approximation at $B=0$. Note that the electron system in high-mobility GaAs
  structures for the typical electron density $n=3\times 10^{11}\,{\rm
  cm}^{-2}$ is only moderately correlated, with the Wigner-Seitz parameter
  $r_s=(\pi na_B^2)^{-1/2}\simeq 1$.}\BibitemShut {Stop}%
\bibitem [{siz()}]{size_eff}%
  \BibitemOpen
  \href@noop {} {}\bibinfo {note} {The observation of a dependence of the
  magnetoresistance on the sample width varying along the sample between
  $100\,\mu{\rm m}$ and $400\,\mu{\rm m}$ in Ref.~\cite{mani13} and between
  $25\,\mu{\rm m}$ and $400\,\mu{\rm m}$ in Ref.~\cite{wang22} raises the
  question as to the spatial distribution of the current density inside the
  sample and, possibly, the character of electron scattering on the edges, but
  the strong magnetoresistance observed in Ref.~\cite{shi14a} for the Hall bar
  width $200\,\mu{\rm m}$ is likely a bulk phenomenon.}\BibitemShut {Stop}%
\bibitem [{rem()}]{remark_mf}%
  \BibitemOpen
  \href@noop {} {}\bibinfo {note} {A useful complementary perspective to
  establish the limits of applicability of the mean-field approximation in the
  2D case can be formulated in the spirit of the framework
  \cite{lanlif8,herring60,stroud76} for describing transport in media with
  macroscopic random inhomogeneities. The ``macroscopic" here means that the
  spatial scale of the inhomogeneities is much larger than the total-momentum
  relaxation length. Our problem is a borderline case for this approach---since
  the total-momentum relaxation length and the length on which the correlations
  of hydrodynamic variables decay because of disorder averaging are both given
  by $L_v$.}\BibitemShut {Stop}%
\end{thebibliography}

\end{document}